\documentclass[graybox]{svmult}

\usepackage{amsmath}
\usepackage{amssymb}
\usepackage{bm}

\usepackage{mathptmx}       
\usepackage{helvet}         
\usepackage{courier}        
\usepackage{type1cm}        
\usepackage{makeidx}         
\usepackage{graphicx}        
\usepackage{multicol}        
\usepackage[bottom]{footmisc}
\usepackage{hyperref}        
\usepackage{soul}            
\hypersetup{colorlinks=true,urlcolor=blue}
\usepackage[square,numbers]{natbib}
\makeindex             

\begin{document}
\title*{Introduction to gravitational wave astronomy}
\author{Nigel T. Bishop \thanks{corresponding author}}
\institute{Nigel T. Bishop \at Department of Mathematics, Rhodes University, Grahamstown 6140, South Africa.\\ \email{n.bishop@ru.ac.za}
}
%
%
\maketitle
\abstract{
This chapter provides an overview of gravitational wave (GW) astronomy, providing background material that underpins the other, more specialized chapters in this handbook. It starts with a brief historical review of the development of GW astronomy, from Einstein's prediction of GWs in 1916 to the first direct detection in 2015. 
It presents the theory of linearized perturbations about Minkowski spacetime of Einstein's equations, and shows how gauge transformations reduce the problem to the standard wave equation with two degrees of freedom, or polarizations, $h_+,h_\times$. We derive the quadrupole formula, which relates the motion of matter in a source region to the far GW field. It is shown that GWs carry energy, as well as linear and angular momentum, away from a source. The GW field of an orbiting circular binary is found; and properties of the evolution of the binary including rate of inspiral and time to coalescence, are calculated. A brief review is given of existing and proposed GW detectors, and of how to estimate source parameters in LIGO or Virgo data of a GW event. The contributions that GW observations have already made to physics, astrophysics and cosmology are discussed.
}
\section*{Keywords} 
Gravitational waves; Quadrupole formula; Orbital inspiral; LIGO; LISA
\section{\textit{Introduction}}

\subsection{\textit{Historical development}}

Soon after proposing the general theory of relativity (GR), Einstein showed, by linearizing the field equations, that the theory implies the existence of gravitational waves (GWs)~\cite{Einstein1916,Einstein1918}, and obtained what has become known as the quadrupole formula.\index{Quadrupole formula} However, the concept of a GW actually predates GR. Newtonian gravitation theory can be expressed as an elliptic equation with the gravitational field changing instantaneously if the source changes. Since special relativity had established the speed of light as a universal speed limit, this suggests that the gravitational theory should be expressed as a hyperbolic system, so implying the existence of GWs.
During the early years, i.e. from about 1920 to 1960, there was uncertainty about the nature of GWs. It was not clear whether GWs carry energy since GW energy at a given event in spacetime cannot be defined; it was not clear whether the quadrupole formul could be applied to gravitationally bound systems; and it was argued that the full nonlinear theory of GR does not permit GW solutions. For a detailed discussion of these issues, see~\cite{Kennefick:1997kb}.

The mathematical theory of GWs became well-established in the 1960s. The above issues were resolved by expressing the Einstein equations of the full nonlinear theory of GR in a suitable coordinate system~\cite{Bondi62,Sachs62}, and also by the ``shortwave approximation'' averaging procedure~\cite{Isaacson68}. Once it was clear that the emission of GWs causes a loss of energy in the emitting system, the dynamics of an orbiting binary and the resulting inspiral could be calculated using linearized theory and the quadrupole approximation~\cite{Peters:1963ux,Peters:1964}.\index{Orbiting binary}

In this period, there were also important developments in the experimental area. The theory of stellar evolution had long indicated that the end-state of a massive star would be a neutron star or a black hole. However, the first observation of a neutron star was of a pulsar in the Crab nebula in 1969, and the first evidence of a massive ($\approx 15M_\odot$) compact object that could only be a black hole was Cygnus X-1 in 1971. The confirmation of the existence of these compact objects was important: their inspiral and merger would be both powerful and in the frequency range $\approx 10$Hz to $\approx 1000$Hz which would be suitable for a terrestrial detector. The first attempt to detect gravitational waves was reported in 1969 using a bar detector~\cite{Weber:1969bz}\index{Bar-detector}. This paper actually reported the detection of a number of GW events, although subsequently it became clear that these events were not astrophysical but rather the result of experimental error. In 1979, observations of the binary pulsar system PSR 1913+16 showed that the orbit was inspiralling at a rate consistent with GW emission in GR, thus providing the first experimental evidence for the existence of GWs~\cite{Taylor79}.\index{Orbiting binary}

The most powerful GW sources, and therefore those most likley to be detected, are the merger of two compact objects, but this case does not satisfy the conditions of the quadrupole approximation and there is a need to go beyond linearized theory. This can be achieved by means of a series expansion, adding terms of quadratic, cubic and higher orders to the linearized expressions. This approach was considered in the early years of GR~\cite{Lorentz17,Lorentz37}, and in the 1980s formalized as the Post-Newtonian method.\index{Post-Newtonian} For the actual merger, a numerical simulation of the full Einstein equations is needed, normally with the spacetime foliated into a sequence of spacelike hypersurfaces~ \cite{Arnowitt62}. In 1977, the GWs from a head-on collision of two Schwarzschild black holes were computed~ \cite{Smarr77}, but it was only in 2005 that codes were able to make a stable evolution of the physically realistic problem of the inspiral and merger of two black holes~ \cite{Pretorius:2005gq}.\index{Numerical relativity} Since then, many other groups have successfully evolved black hole spacetimes, as well as neutron star mergers and supernovae, often using a combination of Post-Newtonian and numerical methods.
  
The possibility of using laser interferometry to detect GWs was suggested in the 1960s, and simple prototypes were constructed at that time.\index{Laser interferometer} In 1980, the US National Science Foundation provided funding for the construction of certain prototypes, as well as for a study of the technical issues and the costs of building an interferometer with arms several km in length. This eventually led to the construction of LIGO (Laser Interferometry Gravitational-Wave Observatory)\index{LIGO} facilities at Hanford and Livingstone, USA, with 4km arms; scientific studies commenced in 2002. Also at this time, the much smaller detector GEO600 (600m arms)\index{GEO600} started operation, and in 2007 Virgo\index{Virgo} in Italy, which has 3km arms and is a project of the European Gravitational Observatory consortium, made its first science run. These instruments underwent a number of upgrades to improve the sensitivity, and the first direct detection of GWs was made on 14 September 2015 by LIGO Hanford and Livingstone~\cite{Abbott2016a}.
 

\subsection{\textit{Content of this chapter}}

This chapter provides an overview of GW astronomy, and describes aspects of the basic theory of GWs that form the foundation of the field. It provides the background material that underpins the specialized chapters in this handbook. The discussion is mainly within the context of the Einstein equations linearized about Minkowski spacetime. This theory is sufficiently straightforward that it can be presented within a single chapter, yet is also widely applicable. The magnitude of GWs is dimensionless, and those propagating past the Earth that have been detected can be characterized as being ${\mathcal O}(10^{-22})$; thus linearized theory certainly applies. The later phase of compact object inspiral is driven by GW emission; this process lasts millions or billions of years and, apart from the final seconds or minutes, is accurately described by linearized theory. Even though linearized theory does not provide an accurate waveform for the actual merger of two compact objects, it remains useful as it provides a simple guide in analytic form.

The theory of linearizing Einstein's equations about Minkowski spacetime is developed in Sec.~\ref{s-Prop}. A key issue is the use of gauge transformations (which, in this context, are coordinate transformations that are almost the identity transformation) to simplify the equations to a form that is manifestly the wave equation. The simplest solution to the wave equation is a plane wave, i.e. in Cartesian coordinates $(t,x,y,z)$ a function $f(t,z)$ that satisfies the wave equation; this solution well describes GWs in the solar system produced by a source many Mpc away. We construct the plane wave solution, and make further gauge transformations to the Transverse Traceless, or TT, gauge. In this way, the 10 components of a symmetric tensor are reduced to 2 independent components or polarizations, usually denoted as $h_+,h_\times$. Finally, this section investigates the effect of a plane GW on a system of test particles. In the TT gauge, the coordinates of the particles do not change, but that does not mean that there is no movement: a coordinate independent quantity such as the proper distance between two of the particles does vary with time. It is also shown that the Riemann tensor has non-zero components, so the spacetime is not flat.

The generation of GWs by the motion of matter is discussed in Sec.~\ref{s-Gen}, and the quadrupole formula is derived. As an example, the formula is applied to the astrophysically important problem of a binary comprising two masses in circular orbit around each other. The two polarization modes, $h_+,h_\times$, are evaluated with respect to spherical polar coordinates. We next discuss the energy, as well as the linear and angular momenta, carried away by GWs. These effects are beyond the scope of linearized theory, being quadratic in the perturbations, and the calculations are only outlined with much detail omitted. This section also summarizes methods that are used for situations where the quadrupole formula is inadequate: Post-Newtonian approximations, numerical relativity, the theory of quasinormal modes of a black hole, and Extreme Mass Ratio Inspirals (EMRIs).

Now that we know what GWs carry away from a system, conservation laws are used in Sec.~\ref{s-OrbBin} to find the evolution of an orbiting circular binary. The orbital diameter and wave period slowly decrease, and we find an expression for the time to coalescence. We also find a relation between the chirp mass, which is a function of the individual masses of the binary, and an expression involving the wave frequency and its rate of change; this means that the chirp mass is entirely determined by observational GW data. It is also shown that GW emission causes an eccentric orbit to circularize, so the focus on circular orbits is for reasons of physics rather than for mathematical convenience.

Sec.~\ref{s-Det} provides an outline of various existing and planned GW detection facilities, including terrestrial laser interferometers, satellite systems, and pulsar timing. We indicate the frequency range in which a detector is sensitive, and some actual or expected astrophysical sources. Then, for the LIGO and Virgo network, we outline the process of determining basic source parameters of an observed GW event representing a compact object inspiral and merger. Some contributions that GW observations have already made to physics, cosmology and astrophysics are outlined.

The chapter ends with a Conclusion, Sec.~\ref{s-Conc}.

\section{\textit{Propagation of gravitational waves}}
\label{s-Prop}

\subsection{\textit{Linearized Einstein equations in vacuum}}
\index{Linearized Einstein equations}
The essential idea is to consider spacetimes that comprise small perturbations about Minkowski spacetime. More precisely, the metric tensor $g_{\alpha\beta}$ is written
\begin{equation}
g_{\alpha\beta}=\eta_{\alpha\beta}+ h_{\alpha\beta}\,,
\label{e-lm}
\end{equation}
where $\eta_{\alpha\beta}=$ is the metric of Minkowski spacetime which in Cartesian $(t,x,y,z)$ coordinates is diag$(-1,1,1,1)$, and $h_{\alpha\beta}$ is a small perturbation. In the linearized approximation, terms of order ${\mathcal O}\left(\left(h_{\alpha\beta}\right)^2\right)$ are neglected. The contravariant metric is written as $g^{\alpha\beta}=\eta^{\alpha\beta}- h^{\alpha\beta}$, and the identities $\delta^\alpha_\gamma=g^{\alpha\beta}g_{\beta\gamma}=\eta^{\alpha\beta}\eta_{\beta\gamma}$ imply that
\begin{equation}
h^{\alpha\beta}=\eta^{\alpha\gamma}\eta^{\beta\delta}h_{\gamma\delta}\,.
\end{equation}
Thus, the indices of quantities of order ${\mathcal O}(h_{\alpha\beta})$ are raised and lowered using the background metric $\eta_{\alpha\beta}$ rather than the full metric $g_{\alpha\beta}$. The first step towards constructing the Einstein equations is to determine the metric connection
\begin{equation}
\Gamma^{\mu}_{\ \alpha\beta} = 
\frac{1}{2} 
\eta^{\mu \nu}(
\partial_{\beta}h_{\nu \alpha} + 
\partial_{\alpha}h_{\nu \beta} - 
\partial_{\nu}h_{\alpha\beta}) =
\frac{1}{2}
(
 \partial_{\beta}h^{\mu}_{\ \ \alpha} + 
 \partial_{\alpha}h^{\mu}_{\ \ \beta} - 
 \partial^{\mu}h_{\alpha \beta}) \,.
 \label{e-Gamma}
\end{equation}
Then the Ricci tensor is
\begin{equation}
\label{lin_rt}
R_{\mu\nu} = 
\partial_{\alpha}\Gamma^{\alpha}_{\ \mu \nu} - 
\partial_{\nu}\Gamma^{\alpha}_{\ \mu \alpha}
= \frac{1}{2}
(
\partial_{\alpha}\partial_{\nu} h^{\ \;\alpha}_{\mu} + 
\partial_{\alpha}\partial_{\mu} h^{\ \;\alpha}_{\nu} -
\partial_{\alpha}\partial^{\alpha} h_{\mu \nu} - 
\partial_{\mu}\partial_{\nu} h
) \,,
\end{equation}
where $h=h^\alpha_\alpha=\eta^{\alpha\beta}h_{\alpha\beta}$. The Ricci scalar is thus
\begin{equation}
\label{lin_rs}
R=R^\mu_{\mu} = 
\partial_{\alpha}\partial_{\beta} h^{\alpha\beta} - 
\partial_{\alpha}\partial^{\alpha} h
 \,,
\end{equation}
and therefore Einstein's equations $G_{\mu\nu}=R_{\mu\nu}-Rg_{\mu \nu}/2=8\pi T_{\mu \nu}$ are
\begin{align}
\label{efe2}
\frac 12 &\left(\partial^{\alpha}\partial_{\nu}   h_{\mu \alpha} + 
\partial^{\alpha}\partial_{\mu}   h_{\nu \alpha} - 
\partial_{\alpha}\partial^{\alpha} h_{\mu \nu} - 
\partial_{\mu}\partial_{\nu}h - 
\eta_{\mu \nu}(
\partial^{\alpha}\partial^{\beta}h_{\alpha\beta} - 
\partial^{\alpha}\partial_{\alpha}h)\right)\nonumber \\
	&= 8 \pi T_{\mu \nu} \,.
\end{align}
Eq.~(\ref{efe2}) may be somewhat simplified by changing the metric perturbation $h_{\alpha\beta}$ to its trace reversed form\index{Trace reversed}, that is
\begin{equation}
\bar{h}_{\alpha\beta}=h_{\alpha\beta}-\frac 1 2 \eta_{\alpha\beta} h
\mbox{  so that  } \bar{h}=-h \mbox{  and  } 
{h}_{\alpha\beta}=\bar{h}_{\alpha\beta}-\frac 1 2 \eta_{\alpha\beta} \bar{h}\,.
\end{equation}
Einstein's equations are then
\begin{equation}
G_{\mu\nu}=-\frac 1 2\left(
\partial_\alpha\partial^\alpha \bar{h}_{\mu\nu} + \eta_{\mu\nu}\partial^\alpha\partial^\beta\bar{h}_{\alpha\beta}
-\partial_\nu\partial^\alpha \bar{h}_{\mu\alpha}-\partial_\mu\partial^\alpha \bar{h}_{\nu\alpha}\right)=8\pi T_{\mu\nu}.
\label{efe3}
\end{equation}
The first term $\partial_\alpha\partial^\alpha \bar{h}_{\mu\nu}$ is the wave operator applied to $\bar{h}_{\mu\nu}$, and the problem would reduce to a standard wave equation if the other terms could be made to disappear. This can be achieved on applying the Lorentz gauge condition
\begin{equation}
\partial^\alpha \bar{h}_{\alpha\beta}=0\,,
\end{equation}
in Eq.~(\ref{efe3}) to obtain
\begin{equation}
-\frac 12 \Box \bar{h}_{\mu\nu}=8\pi T_{\mu\nu}\,,
\label{e-efe4}
\end{equation}
where the operator $\Box = \partial_\alpha\partial^\alpha$, and clearly in vacuum $\Box \bar{h}_{\mu\nu}=0$. Imposition of the Lorentz gauge condition is a constraint on the coordinates being used, and is not a restriction on the geometry of the spacetime. This matter is discussed further in the next section.\index{Lorentz gauge}

\subsection{\textit{Gauge transformations}}
\index{Gauge transformation}
A gauge transformation is a coordinate transformation of the form
\begin{equation}
x^{\mbox{{\tiny (NEW)}}\alpha}=x^{\mbox{{\tiny(OLD)}}\alpha}+\xi^\alpha (x^{\mbox{{\tiny(OLD)}}\beta})
\label{e-gauge}
\end{equation}
where $\partial_\beta\xi^\alpha$ is of the same order of smallness as $h_{\alpha\beta}$; terms of order ${\mathcal O}(h_{\alpha\beta})^2$, ${\mathcal O}(h_{\alpha\beta}\partial_\beta\xi^\alpha)$, ${\mathcal O}(\partial_\beta\xi^\alpha)^2$ are neglected. Applying the coordinate transformation (\ref{e-gauge}) to the metric (\ref{e-lm}) gives
\begin{equation}
h^{\mbox{{\tiny (NEW)}}}_{\alpha\beta}=h^{\mbox{{\tiny (OLD)}}}_{\alpha\beta}
-\partial_\beta\xi_{\alpha}-\partial_\alpha\xi_{\beta}\,,
\end{equation}
and then the change to trace-reversed form leads to
\begin{equation}
\bar{h}^{\mbox{{\tiny (NEW)}}}_{\alpha\beta}=\bar{h}^{\mbox{{\tiny (OLD)}}}_{\alpha\beta}
-\partial_\beta\xi_{\alpha}-\partial_\alpha\xi_{\beta}+\eta_{\alpha\beta}\partial_\mu\xi^\mu\,.
\label{e-hbo2hbn}
\end{equation}
Requiring $\bar{h}^{\mbox{{\tiny (NEW)}}}_{\alpha\beta}$ to satisfy the Lorentz gauge condition then implies\index{Lorentz gauge}
\begin{equation}
\Box \xi_\beta=\partial^\alpha\bar{h}^{\mbox{{\tiny (OLD)}}}_{\alpha\beta}\,.
\label{e-Lxi}
\end{equation}
The right hand side of Eq.~(\ref{e-Lxi}) is regarded as given, and then Eq.~(\ref{e-Lxi}) comprises separate wave equations for each of the four unknowns $\xi_0,\cdots,\xi_3$. The existence of solutions to these four wave equations follows from the theory of partial differential equations. Note however that the solutions are not unique, and we are free to write
\begin{equation}
\xi^\alpha=\xi^{(0)\alpha}+\zeta^\alpha\,,
\end{equation}
where $\xi^{(0)\alpha}$ is a solution to Eq.~(\ref{e-Lxi}), and where $\zeta^\alpha$ satisfies $\Box\zeta^\alpha=0$.

The possibility to make gauge transformations means that a metric that appears to be wavelike may not represent a GW. For example, suppose that the spacetime is Minkowski (and so does not contain gravitational waves) with $\bar{h}^{\mbox{{\tiny (OLD)}}}_{\alpha\beta}=0$, and that $\xi_\alpha=(0,0,0,\epsilon\cos(t-x))$. Then
\begin{equation}
\bar{h}_{\alpha\beta}=\left(
\begin{array}{cccc}
0 & 0 & 0 & \epsilon\sin(t-x) \\
0 & 0 & 0 & -\epsilon\sin(t-x) \\
0 & 0 & 0 & 0 \\
\epsilon\sin(t-x) & -\epsilon\sin(t-x) &0 & 0
\end{array}
\right)\,.
\end{equation}
A simple test to determine whether or not a vacuum spacetime is Minkowski in unusual coordinates, is to evaluate the Riemann tensor: $R^\alpha_{\phantom{\alpha}\beta\gamma\delta}=0$ if and only if the spacetime is  Minkowskian.

\subsection{\textit{Plane wave solutions}}
\label{s-pws}
\index{Plane wave solution}
Let us now make the ansatz
\begin{equation}
\bar{h}^{\alpha\beta}=\Re\left( A^{\alpha\beta}\exp(i k_\alpha x^\alpha)\right)\,,
\label{e-h2A}
\end{equation}
where the $k_\alpha$ are real constants and the $A_{\alpha\beta}$ are complex constants; note that a general wave may be expressed as a Fourier sum of terms with the above form. Substituting into the vacuum field equations $\Box\bar{h}_{\alpha\beta}=0$ leads to $\eta^{\alpha\beta}k_\alpha k_\beta=0$, so that $k_\alpha$ is a null vector, and usually $k_0$ is written as $\omega$. Since $A^{\alpha\beta}$ is symmetric, it has 10 independent components. Now, the Lorentz gauge condition $\partial^\beta\bar{h}_{\alpha\beta}=0$ implies 4 conditions\index{Lorentz gauge}
\begin{equation}
A^{\alpha\beta}k_\beta=0\,,
\label{e-Ak}
\end{equation}
reducing the number of independent components of $A^{\alpha\beta}$ from 10 to 6. Then
a gauge transformation can be made where
\begin{equation}
\xi^\alpha=\Re(B^\alpha\exp(ik_\mu x^\mu))
\label{e-gB}
\end{equation}
with $B^\alpha$ constant; since $\xi^\alpha$ satisfies the identity $\Box\xi^{\alpha}=0$, the condition $\partial_\beta\bar{h}^{\alpha\beta}=0$ is not affected. The 4 constants $B^\alpha$ are used to reduce the independent components of $A^{\alpha\beta}$ from 6 to 2.\index{Gauge transformation}

In order to apply the above and construct in a transparent way an explicit form of $A^{\alpha\beta}$ with only two independent components, we align the coordinates with the direction of propagation of the wave. Let the wave be propagating in the $z-$direction with frequency $\omega/(2\pi)$ so that $k_\alpha=(\omega,0,0,-\omega)$. Then the conditions~(\ref{e-Ak}) imply that $A^{\alpha 0}=A^{\alpha 3}$, so we can write
\begin{equation}
A^{\alpha\beta}=\left(
\begin{array}{cccc}
A^{00} & A^{01} & A^{02} & A^{00} \\
A^{01} & A^{11} & A^{12} & A^{01} \\
A^{02} & A^{12} & A^{22} & A^{02} \\
A^{00} & A^{01} & A^{02} & A^{00}
\end{array}
\right)\,,
\end{equation}
which has six independent components $A^{00}, A^{01}, A^{02},A^{11}, A^{12}, A^{22}$. Differentiating the gauge transformation~(\ref{e-gB}) gives $\partial^\alpha\zeta^\beta=\Re(iB^\alpha k^\beta\exp(ik_\mu x^\mu))$, so that from Eq.~(\ref{e-hbo2hbn}),
\begin{equation}
\bar{h}^{\prime\alpha\beta}=\Re\left( (A^{\alpha\beta}-iB^\alpha k^\beta-iB^\beta k^\alpha+\eta^{\alpha\beta}i B^\mu k_\mu) \exp(i k_\mu x^\mu)\right)\,.
\end{equation}
Now, $A^{\alpha\beta},B^\alpha,(k^{\prime\alpha}-k^\alpha),(x^{\prime\alpha}-x^\alpha)$ are small quantities, and so the product of any pair is negligible. Thus we can write $\bar{h}^{\prime\alpha\beta}=\Re\left( (A^{\prime\alpha\beta}\exp(i k_\mu x^\mu)\right)$, and then
\begin{equation}
A^{\prime\alpha\beta}=A^{\alpha\beta}-iB^\alpha k^\beta-iB^\beta k^\alpha+\eta^{\alpha\beta}i B^\mu k_\mu\,.
\label{e-A2Aprime}
\end{equation}
\begin{align}
A^{\prime 00}&=A^{00}+i\omega(B^0+B^3)\,,\qquad A^{\prime 01}=A^{01}+i\omega B^1\,,
\qquad A^{\prime 02}=A^{02}+i\omega B^2
\nonumber\\
A^{\prime 11}&=A^{11}+i\omega(B^3-B^0)\,,\qquad
A^{\prime 22}=A^{22}+i\omega(B^3-B^0)\,,\qquad
A^{\prime 12}=A^{12}\,.
\end{align}
The constants $B^\alpha$ may be chosen so as to simplify $A^{\prime\alpha\beta}$ in any desired way, and common practice is to use them to set
\begin{equation}
A^{\prime 00}=A^{\prime 01}=A^{\prime 03}=0\,,\;\;A^{\prime 22}=-A^{\prime 11}\,,
\end{equation}
so that $A^{\prime \alpha\beta}$ is
\begin{equation}
A^{\prime\alpha\beta}=\left(
\begin{array}{cccc}
0 & 0 & 0 & 0 \\
0 & A^{\prime 11} & A^{\prime 12} & 0 \\
0 & A^{\prime 12} & -A^{\prime 11} & 0 \\
0 & 0 & 0 & 0
\end{array}
\right)\,.
\label{e-A-TT}
\end{equation}

\subsection{\textit{The TT gauge}}
\index{TT gauge}
The resulting metric $\bar{h}_{\alpha\beta}$ is {\bf transverse}: the wave travels in the $z$-direction, and its effects are in the $x$- and $y$-directions. More generally, the transverse property is expressed as
\begin{equation}
\bar{h}^\prime_{\alpha\beta} k^\alpha=0\,.
\label{e-transverse}
\end{equation}
Further the metric is {\bf traceless}, i.e.
\begin{equation}
\bar{h}^{\prime\alpha}_\alpha=\bar{h}^\prime=0\,,
\label{e-traceless}
\end{equation}
so that the metric and its trace-reverse form are the same, $\bar{h}^\prime_{\alpha\beta}=h^\prime_{\alpha\beta}$. A metric satisfying Eqs.~(\ref{e-transverse}) and (\ref{e-traceless}) is said to be in the {\bf Transverse Traceless} or {\bf TT} gauge. From now on, we will drop the ${}^\prime$ to denote a quantity in this gauge, and in circumstances where there is a need to differentiate the gauge of quantities, we will use the superfix ${}^{TT}$ to denote a quantity in the TT gauge.

The GW derived above has two independent components $h^{11},h^{12}$. However, there is no essential difference between the components, as can be seen by making a coordinate transformation which, geometrically, corresponds to a rotation of the $(x,y)$ axes about the $z$-axis through an angle $\theta=\pi/4$. More precisely, $(t,x,y,z)\rightarrow ((t^\prime,x^\prime,y^\prime,z^\prime)$ where
\begin{equation}
x^\prime=x\cos\theta+y\sin\theta\,,\qquad y^\prime=x\sin\theta-y\cos\theta\,,
\label{e-rotate}
\end{equation}
and here $\cos\theta=\sin\theta=\sqrt{2}/2$. Application to the metric $g^{\alpha\beta}=\eta^{\alpha\beta}-h^{\alpha\beta}$ leads to
\begin{equation}
h^{\prime 11}=-h^{\prime 22}=h^{12}\,,\qquad h^{\prime 12}=h^{\prime 21}=-h^{11}\,.
\end{equation}

\index{Polarization modes $h_+,h_\times$}
The two independent components of a GW, $h^{11},h^{12}$, are usually denoted by $h_+,h_\times$ respectively and are called the two polarization states of the GW. There is an analogy to electromagnetic waves which are also transverse to the direction of propagation and which also have polarization states, although in this case the angle between the two states is $\pi/2$ rather than $\pi/4$. As shown above, the distinction between $h_+,h_\times$ is observer dependent, but in some circumstances the polarization can be described in an observer-independent way. We write
\begin{equation}
h_+=\Re(A_+\exp\left(i\omega(t-z))\right)\,,\;\;
h_\times=\Re(A_\times\exp\left(i\omega(t-z))\right)\,.
\label{e-h+A+}
\end{equation}
Now, $A_+,A_\times$ are complex constants (i.e., independent of the coordinates $x^\alpha$), and consider the ratio $A_\times/A_+=R_A$ which must also be constant. There are two interesting special cases: (1) Suppose that the ratio $R_A$ has zero imaginary part, then $h_\times/h_+$ is constant, and so we may change coordinates by rotating the $(x,y)$ axes about the $z$-axis through some angle $\theta$ to achieve $h_\times=0$; in this case the GW is said to be {\bf linearly polarised}. (2) Suppose that $R_A=\pm i$, then from the identity $\cos^2s+\sin^2s=1$, it follows that the magnitude of the GW $\sqrt{h_+^2+h_\times^2}$ is constant, and the wave is said to be {\bf circularly polarised}. The physical properties leading to these names are discussed in the next sub-section.

\subsection{\textit{The effect of a gravitational wave on a system of test particles}}
\index{TT gauge}
We now consider the physical effects of a GW in the TT gauge as described in the previous sub-section. Since $h_{\alpha 0}=0$, it follows from Eq.~(\ref{e-Gamma}) that $\Gamma^\alpha_{00}=0$. The geodesic equation is
\begin{equation}
\frac{d^2x^\alpha}{d\tau^2}=-\Gamma^\alpha_{\mu\nu}\frac{dx^\mu}{d\tau}\frac{dx^\nu}{d\tau}\,,
\end{equation}
where $\tau$ is the proper time; if a particle is initially at rest, ${dx^\mu}/{d\tau}=(1,0,0,0)$ so that ${d^2x^\alpha/}{d\tau^2}=0$ and the particle remains at rest. But this is just a consequence of the coordinates being used, and does not mean that the GW has no physical effect. In order to analyze this further, we use the equation of geodesic deviation. Consider two nearby particles, $A$ and $B$, each with 4-velocity $U^\alpha$, and let $c^\alpha$ be a vector connecting $A$ and $B$ and orthogonal to $U^\alpha$; then
\begin{equation}
\frac{d^2 c^\alpha}{d\tau^2}+\Gamma^\alpha_{\beta\gamma}U^\beta U^\gamma
=-R^\alpha_{\phantom{\alpha}\beta\gamma\delta} U^\beta U^\delta c^\gamma\,.
\label{e-geod-dev}
\end{equation}
Now suppose that we use proper coordinates with $A$ located at the origin; then $U^\alpha=(1,0,0,0)$, $c^\alpha=(0,c^i)=(0,x_B^i)$ (with $i=1,2,3$ denoting a spatial index), the metric is
\begin{equation}
ds^2=\left(\eta_{\alpha\beta}+{\mathcal O}(|x^i|^2)\right)dx^\alpha dx^\beta\,,
\end{equation}
$\Gamma^\alpha_{\beta\gamma}=0$, $\tau=t$, and Eq.~(\ref{e-geod-dev}) simplifies to
\begin{equation}
\frac{d^2 x_B^i}{dt^2}=-R^i_{\phantom{i}0j 0}  x_B^j\,.
\label{e-geod-dev2}
\end{equation}

It is important to note that $R^\alpha_{\phantom{\alpha}\beta\gamma\delta}$ is {\bf gauge-invariant},\index{Gauge-invariant} that is if we make a gauge transformation given by Eq.~(\ref{e-gauge}), then in the new gauge $R^{\mbox{{\tiny (NEW)}}\alpha}_{\phantom{\mbox{{\tiny (NEW)}}\alpha}\beta\gamma\delta}=R^{\mbox{{\tiny (OLD)}}\alpha}_{\phantom{\mbox{{\tiny (OLD)}}\alpha}\beta\gamma\delta}$. This result follows since $R^\alpha_{\phantom{\alpha}\beta\gamma\delta}={\mathcal O}(h_{\alpha\beta})$, and the coordinate transformation matrix is
\begin{equation}
\frac{\partial x^{\mbox{{\tiny (NEW)}}\alpha}}{\partial x^{\mbox{{\tiny(OLD)}}\beta}}
=\delta^\alpha_\beta+{\mathcal O}(h_{\alpha\beta})\,,
\end{equation}
so to ${\mathcal O}(h_{\alpha\beta})$ only the $\delta^\alpha_\beta$ is used in the coordinate transformation. Thus we may use the TT gauge to evaluate the Riemann tensor in Eq.~(\ref{e-geod-dev2}), and we find
\begin{equation}
R^1_{\phantom{1}010}=-R^2_{\phantom{2}020}=-\frac 12 \partial^2_t h_+\,,\;\;
R^1_{\phantom{1}020}=R^2_{\phantom{2}010}=-\frac 12 \partial^2_t h_\times\,,
\end{equation}
with the remaining components of the form $R^i_{\phantom{i}0j0}$ being zero. Then Eq.~(\ref{e-geod-dev2}) simplifies to
\begin{equation}
\frac{d^2 x_B^i}{dt^2}=\left(\frac 12 \partial^2_t h_+ x_B^1
+\frac 12 \partial^2_t h_\times x_B^2,
\frac 12 \partial^2_t h_\times x_B^1-\frac 12 \partial^2_t h_+ x_B^2,
0\right)\,.
\label{e-geod-dev3}
\end{equation}
We now write $x^i_B(t)=x^i_{B0}+\epsilon^i(t)$ with $\epsilon^i$ of order ${\mathcal O}(h)$, and replace $x^i_B$ by $x^i_{B0}$ on the right hand side of Eq.~(\ref{e-geod-dev3}) (as terms of order ${\mathcal O}(h^2)$ are ignorable) so that Eq.~(\ref{e-geod-dev3}) can be integrated to give
\begin{equation}
\epsilon^i=\left(\frac 12 h_+ x_{B0}^1
+\frac 12 h_\times x_{B0}^2,
\frac 12 h_\times x_{B0}^1-\frac 12 h_+ x_{B0}^2,
0\right)\,.
\label{e-geod-dev4}
\end{equation}

\begin{figure}
\begin{center}
\includegraphics[width=12cm]{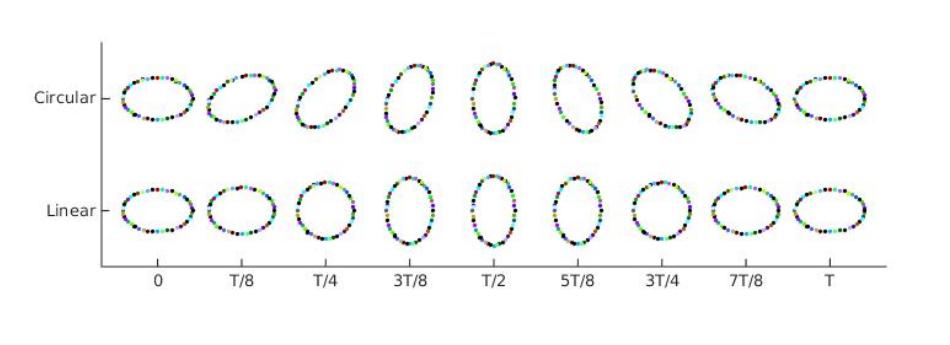}
\end{center}
\caption{The effect on a circle of particles transverse to a GW. Upper row: circular polarization, Lower row: linear polarization. In the horizontal axis, $T$ is the period of the GW. The particles are in their unperturbed state at $T/4$ and $3T/4$ in the lower row. In the circular polarization case it may appear that the whole ellipse is rotating, but actually particles are moved a little from the unperturbed state to give this impression; in the same way that a wave moving across the ocean is caused by water moving up and down, not horizontally.}\index{Polarization modes $h_+,h_\times$}
\label{f-polarize}
\end{figure}

In order to facilitate the interpretation of Eq.~(\ref{e-geod-dev4}), let us suppose that the $(x^1,x^2)$ axes have been rotated so that $x^2_{B0}=0$, use Eq.~(\ref{e-h+A+}), and suppose that the origin of the spacetime coordinates is such that $A_+$ is real. Then in the case of a linearly polarised wave
\begin{equation}
\epsilon^i=\frac 12 x_{B0}^1 A_+\cos\omega t\left( 1,R_A,0\right)\,,
\end{equation}
so that $B$ moves in a straight line; and in the case of a circularly polarized wave
\begin{equation}
\epsilon^i=\frac 12 x_{B0}^1 A_+\left( \cos\omega t,\sin\omega t,0\right)\,,
\end{equation}
so that $B$ moves in a circle. The effect of a GW, for both linear and circular polarization states is illustrated in Fig.~\ref{f-polarize}.

The preceeding analysis brings out two important general points. Firstly, since $R^\alpha_{\phantom{\alpha}\beta\gamma\delta}\ne 0$, the spacetime is not Minkoskian. Secondly, suppose that the particles $A$ and $B$ are massive and are connected by a damping system; then the relative motion and acceleration will mean that work is done on the damping system and therefore its temperature must increase. These considerations show that GWs are real physical effects that contain energy that must have originated in some source and which can, in principle, be extracted from the GWs.

\section{\textit{Generation of gravitational waves}}
\label{s-Gen}

\subsection{\textit{Linearized Einstein equations with matter: the quadrupole formula}}
\index{Linearized Einstein equations}\index{Quadrupole formula}
From standard wave equation theory, the solution to Eq.~(\ref{e-efe4}) representing an outgoing wave is
\begin{equation}
\bar{h}^{\mu\nu}(t,x^i)=4 \int \frac{ T^{\mu\nu}(t-|x^i-x^{\prime i}|,x^{\prime j})}{|x^i-x^{\prime i}|} d^3 x^\prime\,,
\label{e-quad1}
\end{equation}
where $(t,x^i)$ is the event at which the metric perturbation $\bar{h}^{\mu\nu}$ is evaluated, $x^{\prime i}$ represents a spatial point inside the source, and $|x^i-x^{\prime i}|$ is the Euclidean distance between $x^i$ and $x^{\prime i}$. Suppose now that the source is matter localized near the origin in a region $r=|x^i|<r_0$, and we are evaluating the metric perturbations where $r={\mathcal O}(r_E)$ with $r_0\ll r_E$;  then Eq.~(\ref{e-quad1}) simplifies to
\begin{equation}
\bar{h}^{\mu\nu}(t,x^i)=\frac 4r \int  T^{\mu\nu}(t-r,x^{\prime j}) d^3 x^\prime\,.
\label{e-quad2}
\end{equation}

The discussion in Sec.~\ref{s-pws} showed that the 10 components of $\bar{h}^{\mu\nu}$ have only 6 independent components, so it is sufficient to determine the spatial components $\bar{h}^{ij}$. In the linearized approximation, the conservation condition $\nabla_\alpha T^{0\alpha}=0$ simplifies to $\partial_t T^{00}+\partial_i T^{0i}=0$, and applying this relation twice gives
\begin{equation}
\partial_i\partial_j T^{ij}=\partial^2_tT^{00}\,.
\end{equation}
Using the above together with the rules of basic calculus gives
\begin{align}
\partial^2_t (T^{00}x^ix^j)=&(\partial_k\partial_\ell T^{k\ell})x^ix^j \nonumber \\
=&\partial_k\partial_\ell (T^{k\ell}x^ix^j)
-2\partial_k(T^{ki}x^j+T^{kj}x^i)+2T^{ij}\,,
\end{align}
from which it follows that
\begin{equation}
\int T^{ij} d^3x=\frac 12 \partial^2_t\int T^{00}x^i x^j d^3x\,,
\end{equation}
since the remaining terms $\partial_k\left[\partial_\ell (T^{k\ell}x^ix^j)-2(T^{ki}x^j+T^{kj}x^i)\right]$ can be transformed by the divergence theorem into a surface integral over the boundary of the volume of integration where $T^{\alpha\beta}=0$.

As well as the gravitational field being weak with $h_{\mu\nu}\ll 1$, we now assume that throughout the source region the relative speed of the matter flow is much less than that of light. Then $T^{00}$ can be replaced by the matter density $\rho$ so that Eq.~(\ref{e-quad1}) becomes
\begin{equation}
\bar{h}^{ij}(t,x^i)=\frac 2r \frac{d^2 I^{ij}(t-r)}{dt^2}\,,
\label{e-quad3}
\end{equation}
where the 3-dimensional tensor
\begin{equation}
I^{ij}(t)=\int\rho(t,x^{\prime k}) x^{\prime i}x^{\prime j} d^3x
\label{e-quad4}
\end{equation}
is the second moment, also called the quadrupole moment\index{Quadrupole moment}, of the mass distribution. In the case that the matter distribution is treated as a system of $N$ particles,
\begin{equation}
I^{ij}(t)=\sum_{a=1}^N M_a x^{\prime i}_ax^{\prime j}_a\,.
\label{e-quad5}
\end{equation}

\subsubsection{\textit{Example: Equal mass circular binary}}
\index{Orbiting binary}
Consider two particles, $A$ and $B$, each of mass $M$ in circular orbit around each other with orbital diameter $r_0$, and suppose that the angular velocity of each particle is $\omega$. Choosing coordinates so that the orbit is in the $x-y$ plane, the position vector of particle $A$ is
\begin{equation}
x^i_A=\left(\frac{r_0}{2}\cos\omega t,\frac{r_0}{2}\sin\omega t,0\right)\,,
\end{equation}
and for particle $B$, $x_B^i=-x^i_A$. Then 
\begin{equation}
I^{ij}=
\frac{Mr_0^2}{2}\left(
\begin{array}{ccc}
\cos^2\omega t &\cos\omega t \sin\omega t & 0 \\
\cos\omega t \sin\omega t & \sin^2\omega t & 0 \\
0 & 0& 0
\end{array}
\right)\,.
\end{equation}
Then using Eq.~(\ref{e-quad3}), it follows that
\begin{equation}
\bar{h}^{ij}=
-\frac{2Mr_0^2\omega^2}{r}\left(
\begin{array}{ccc}
\cos 2\omega(t-r) &\sin 2\omega(t-r) & 0 \\
\sin 2\omega(t-r) & -\cos 2\omega(t-r) & 0 \\
0 & 0& 0
\end{array}
\right)\,.
\label{e-emcb}
\end{equation}

\index{TT gauge}
For an observer on the $z$-axis, Eq.~(\ref{e-emcb}) is already in the TT gauge, but for any other observer a transformation to the TT gauge is needed. We now construct the transformation for an observer on the $x$-axis, by amending the theory developed in Sec.~\ref{s-pws}. The wave propagation vector becomes $k_\mu =(2\omega,-2\omega,0,0)$; then using Eq.~(\ref{e-Ak}) and noting the zero entries in Eq.(\ref{e-emcb}), we have
\begin{equation}
A^{\alpha\beta}=\left(
\begin{array}{cccc}
A^{11} & A^{11} & A^{12} & 0 \\
A^{11} & A^{11} & A^{12} & 0 \\
A^{12} & A^{12} & A^{22} & 0 \\
0 & 0 & 0 & 0
\end{array}
\right)\,.
\end{equation}
Eq.~(\ref{e-A2Aprime}) is now applied, giving
\begin{align}
A^{\prime 11}=&A^{11}+i2\omega(B^0+B^1)\,,\qquad A^{\prime 12}=A^{12}+i2\omega B^2\,,
\qquad A^{\prime 13}=i2\omega B^3
\nonumber\\
A^{\prime 22}=&A^{22}+i2\omega(B^1-B^0)\,,\qquad
A^{\prime 23}=0\,,\qquad
A^{\prime 33}=i2\omega(B^1-B^0)\,.
\end{align}
The TT condition $A^{\prime 22}+A^{\prime 33}=0$ is satisfied upon setting $i2\omega(B^1-B^0)=-A^{22}/2$, and then $(B^0+B^1),B^2,B^3$ are chosen so that $A^{\prime 11}=A^{\prime 12}=A^{\prime 13}=0$. Thus, 
\begin{equation}
\bar{h}^{ij}=
\frac{Mr_0^2\omega^2}{r}\left(
\begin{array}{ccc}
0 & 0& 0 \\
0&\cos 2\omega(t-r) &0  \\
0&0& -\cos 2\omega(t-r)  \\
\end{array}
\right)\,.
\end{equation}
Finally, in order to be able to make a consistent comparison of the waveform in different directions, we need to rotate the $(x,z)$ coordinates about the $y$-axis through an angle $\pi/2$ so that the GW is travelling in the $z$-direction. The result is
\begin{equation}
\bar{h}^{ij}=
\frac{Mr_0^2\omega^2}{r}\left(
\begin{array}{ccc}
-\cos 2\omega(t-r) & 0& 0 \\
0&\cos 2\omega(t-r) &0  \\
0&0& 0  \\
\end{array}
\right)\,.
\end{equation}

In the general case, the observer's position is expressed using spherical polars $(r,\theta,\phi)$; without loss of generality, we set $\phi=0$ since a non-zero $\phi$ is equivalent to changing $t$ to $t-\phi/\omega$. The wave propagation vector becomes $k_\mu=(\omega,-\omega\sin\theta,0,-\omega\cos\theta)$, and the transformation to the TT gauge proceeds in a similar way to that described above, but the details are omitted.\index{TT gauge} In order for the TT form to be expilicit, we must also make a coordinate transformation that represents a rotation through an angle $\theta$ of the $x,z$ coordinates about the $y-$axis. The result for a general circular binary with masses $M_1,M_2$ is
\begin{align}
h_+=&-\frac{2\omega^2 \mu r_0^2}{r}(1+\cos^2\theta)\cos(2\omega(t-r)-2\phi)\nonumber \\
h_\times=&-\frac{4\omega^2 \mu r_0^2}{r}\cos\theta\sin(2\omega(t-r)-2\phi)\,,
\label{e-hcircbin}
\end{align}
where $\mu=M_1M_2/(M_1+M_2)$ is $M/2$ for the equal mass case, and $\mu$ is called the reduced mass of the binary. The magnitudes of both GW components $h_+,h_\times$ are maximum at $\theta=0,\pi$ (i.e. for an observer seeing the binary ``face-on''), and minimum for an observer at $\theta=\pi/2$ (i.e., for an observer in the plane of the binary). Note further that at $\theta=0,\pi$ the GW is circularly polarized, whereas at $\theta=\pi/2$ it is linearly polarized.\index{Polarization modes $h_+,h_\times$}

\subsection{\textit{Energy of gravitational waves}}
\index{Energy of gravitational waves}
The energy associated with gravitational waves is a second-order effect, and so its justification goes beyond the linearized approximation considered so far. Here, we outline the calculation, but with many details omitted~\cite{Isaacson68}. Eq.~(\ref{e-lm}) is amended to
\begin{equation}
g_{\alpha\beta}=g^{[B]}_{\alpha\beta}+ h^{[1]}_{\alpha\beta}+h^{[2]}_{\alpha\beta}\,,
\label{e-lm2}
\end{equation}
where: $h^{[1]}_{\alpha\beta}$ is $h_{\alpha\beta}$ in Eq.~(\ref{e-lm}) and is the first-order perturbation; $h^{[2]}_{\alpha\beta}$ is the second-order perturbation; and $g^{[B]}_{\alpha\beta}$ is the background metric, and is a little different to the Minkowski metric $\eta_{\alpha\beta}$ because it includes the response of the background geometry to the first-order perturbations. Evaluation of the Ricci tensor for the metric of Eq.~(\ref{e-lm2}) gives a similar expansion
\begin{equation}
R_{\alpha\beta}=R^{[B]}_{\alpha\beta}(g^{[B]})+ R^{[1]}_{\alpha\beta}(h^{[1]})+R^{[2]}_{\alpha\beta}(h^{[1]})+R^{[1]}_{\alpha\beta}(h^{[2]})\,,
\label{e-lm2R}
\end{equation}
where $R^{[1]}_{\alpha\beta}(h^{[1]})$ is of order ${\mathcal O}\left(h^{[1]}\right)$, and the last two terms are of order ${\mathcal O}\left(\left(h^{[1]}\right)^2\right)$. 
Then solving Eq.~(\ref{e-lm2R}) to first-order in $h^{[1]}$ gives $R^{[1]}_{\alpha\beta}(h^{[1]})=0$, from which the linearized theory discussed in the preceeding sections of this chapter is obtained.

The next step is to split the second-order part of $R_{\alpha\beta}$ into a part that varies on scales much larger than the wavelength of the GWs, and a remainder that contains the local fluctuations. Defining $<x>$ to be the average of the quantity $x$ over a suitable region (in a sense made precise in~\cite{Isaacson68}), then
\begin{align}
R^{[B]}_{\alpha\beta}(g^{[B]})+ <R^{[2]}_{\alpha\beta}(h^{[1]})>&=0\,,
\label{e-lmR2} \\
R^{[1]}_{\alpha\beta}(h^{[2]})+R^{[2]}_{\alpha\beta}(h^{[1]})-
<R^{[2]}_{\alpha\beta}(h^{[1]})>
&=0\,.
\end{align}
We can then define
\begin{equation}
T^{[GW]}_{\alpha\beta}=-\frac{1}{8\pi}\left(<R^{[2]}_{\alpha\beta}(h^{[1]})>
-\frac 12 g^{[B]}_{\alpha\beta}<R^{[2]}(h^{[1]})>\right)\,,
\end{equation}
since, from Eq.~(\ref{e-lmR2}), the Einstein tensor for the background metric can be written as
\begin{equation}
G^{[B]}_{\alpha\beta}=R^{[B]}_{\alpha\beta}-\frac 12 g^{[B]}_{\alpha\beta}R^{[B]}=
-\left(<R^{[2]}_{\alpha\beta}(h^{[1]})>-\frac 12 g^{[B]}_{\alpha\beta}<R^{[2]}(h^{[1]})>\right)\,.
\end{equation}

Evaluation of $T^{[GW]}_{\alpha\beta}$ leads to a simple form when $h^{[1]}_{\alpha\beta}$ satisfies the Lorentz and TT gauge conditions\index{Lorentz gauge}\index{TT gauge}
\begin{equation}
T^{[GW]}_{\alpha\beta}=\frac{1}{32\pi}<(\partial_\alpha h_{\mu\nu})
(\partial_\beta h^{\mu\nu})>\,.
\end{equation}
In the case of GWs moving radially outwards from a source which is far away, we can write
\begin{equation}
T^{[GW]}_{00}=-T^{[GW]}_{01}=T^{[GW]}_{11}=\frac{1}{32\pi}<(\partial_t h_{\mu\nu})
(\partial_t h^{\mu\nu})>=\frac{1}{16\pi}<(\partial_t h_+)^2+
(\partial_t h_\times)^2>\,.
\label{e-TGW}
\end{equation}
The stress-energy tensor $T^{[GW]}_{\alpha\beta}$ can be written in the form $\rho k_\alpha k_\beta$ with $k_\alpha$ being the null vector $(-1,1,0,0)$, and $\rho$ being the right hand side of Eq.~(\ref{e-TGW}); thus $T^{[GW]}_{\alpha\beta}$ can be interpreted as being sourced by dust comprising massless particles moving radially outwards at the speed of light. The total power crossing the 2-surface $S$ at $r=$constant, $t=$ constant is then
\begin{equation}
L^{[GW]}=\int_S T^{[GW]01} r^2 d\Omega= \frac{r^2}{16\pi}\int_S (\partial_t h_+)^2+
(\partial_t h_\times)^2  d\Omega\,,
\label{e-LGW}
\end{equation}
with the averaging $<>$ removed since $h_+,h_\times$ vary slowly in the angular directions. For example, in the case of a circular binary with $h_+,h_\times$ given by Eq.~(\ref{e-hcircbin}),\index{Orbiting binary}
\begin{equation}
L^{[GW]}=\frac{32}{5}\mu^2 r_0^4\omega^6\,.
\label{e-LGW-circbin}
\end{equation}

Eq.~(\ref{e-LGW}) describes a power output, and conservation of energy means that it should be balanced by an energy loss somewhere else. Indeed, there is a precise result to this effect~\cite{Bondi62,Sachs62}, which we briefly outline. The mathematical framework used is coordinates based on outgoing null cones. For example, the metric of Minkowski spacetime is
\begin{equation}
ds^2=-du^2 -2 du\, dr +r^2(d\theta^2+\sin^2\theta d\phi^2)
\label{e-nullM}
\end{equation}
in which the coordinate transformation $t\rightarrow u=t-r$ has been applied to the usual Minkowski metric in spherical polar coordinates ($ds^2=-dt^2+dr^2+r^2(d\theta^2+\sin^2\theta d\phi^2)$). For a general spacetime, the metric is of the form Eq.~(\ref{e-nullM}) plus additional terms (which may not be small). However, the key assumption made is that the geometry is asymptotically flat, which means that the additional terms $\rightarrow 0$ at least as fast as $1/r$ as $r\rightarrow\infty$. In the general case, the Einstein equations are very complicated, but if we consider only the leading-order term (in an asymptotic expansion in $1/r$), then they simplify enormously and it can be shown that
\begin{equation}
\frac{dm}{du}=-L^{[GW]}\,.
\label{e-BondiNews}
\end{equation}
In general, the physical meaning of $m$ is ambiguous, but in the case that the spacetime is static the physical meaning is clear: it is the Schwarzschild mass of the system. We can therefore consider a system that is static until time $u_1$, then emits GWs until time $u_2$ from when it is again static:
\begin{equation}
m(u_2)-m(u_1)=-\int_{u_1}^{u_2} L^{[GW]} du\,.
\end{equation}
If the dimensionless quantity $dm/du\ll 1$ then the system is regarded as quasi-static, and $m$ in Eq.~(\ref{e-BondiNews}) is treated as the Schwarzschild mass. What does it mean to say that the total mass of a system, for example an orbiting binary, is decreasing? In general relativity, all forms of energy contribute to the total mass. If the system is approximately Newtonian and satisfies the conditions for the quadrupole formula to be valid, then the total mass is the sum of the rest masses of the orbiting objects (which does not change), together with the Newtonian kinetic plus potential energies. This matter is pursued further in Sec. \ref{s-inspiral}.\index{Quadrupole formula}

As well as energy, GWs may also carry angular momentum and linear momentum. The formula for the linear momentum $P^i$ is\index{Angular momentum of gravitational waves}\index{Momentum of gravitational waves}
\begin{equation}
\frac{dP^i}{dt}= -\frac{r^2}{16\pi}\int_S n^i\left((\partial_t h_+)^2+
(\partial_t h_\times)^2\right)  d\Omega\,,
\label{e-dPdt}
\end{equation}
where $n^i$ is the unit vector normal to $S$. When the motion of the source is in the $x-y$ plane, then the angular momentum $J_i$ is in the $z$-direction, and is
\begin{equation}
J_z=\frac{r^2}{16\pi}\int_S \partial_t h_+\partial_\phi h_+ +
\partial_t h_\times\partial_\phi h_\times  d\Omega\,.
\label{e-dJzdt}
\end{equation}
These formulas are derived, for example, in~\cite{Alcubierre:2008}. Further, even when the source is not weak, sufficiently far from the source $h_+,h_\times$ are small and so in the limit as $r\rightarrow \infty$ Eqs~(\ref{e-LGW}), (\ref{e-dPdt}), and (\ref{e-dJzdt}) still apply.

\subsection{\textit{Beyond the quadrupole formula}}
The derivation of the quadrupole formula assumed that velocities are small, but that does not apply for compact objects near merger. In this case, an accurate GW calculation requires going beyond the quadrupole formula. The various options available are summarized here.

\index{Quadrupole formula}
The {\bf post-Newtonian (PN)}\index{Post-Newtonian} method gives a formula for the GWs as a series with the first term in the series being the quadrupole result; thus, the method is an extension of the quadrupole formula. The series can be regarded as being in terms of powers of $v$; however, the formalism used in PN work normally uses SI (with $G,c\ne 1$) rather than geometric units, and a term in the series expansion of the form $1/c^{2n}$ is said to be of PN order $n$. The quadrupole term is PN order 1, and other terms are PN order $\frac 32$ or higher. PN methods accurately describe the inspiral of two compact objects; further, results are obtained simply by evaluating analytic formulas, and so the computation of the evolution of the system is very quick. However, as merger approaches the gravitational field becomes highly nonlinear and velocities may approach that of light, and the PN approximation ceases to be accurate. The PN method is discussed further in chapter 22 in this book; see also the review article~\cite{Blanchet06}.

In {\bf numerical relativity}\index{Numerical relativity}, the full nonlinear Einstein equations are solved. The spacetime is foliated into a sequence of 3D hypersurfaces, and data on an initial hypersurface is evolved to the next hypersurface; the process is then repeated until the desired region of the spacetime has been covered. While numerical relativity is accurate even when the fields are highly nonlinear, it can require substantial computational resources. Thus, the GWs emitted by a compact binary system are normally calculated using the PN method during the inspiral, followed by numerical relativity for the final stages of the inspiral and the merger. As with PN methods, numerical relativity is a huge field, but it is not discussed further here. It is covered in chapter 23 in this book, and also in the text-books~\cite{Alcubierre:2008,Baumgarte2010}.\index{Orbiting binary}

{\bf Quasinormal modes}\index{Quasinormal modes} are perturbations on a Kerr background that satisfy regularity conditions at the black hole horizon and infinity. After merger, and assuming a black hole is formed, a system will emit GWs at a frequency and decay rate that depend only on the final mass and angular momentum of the black hole. Details can be found in~\cite{Chandrasekhar98,Kokkotas99a}. This process is called the ``ring-down'', and is quite short compared to the inspiral and merger phases. Physically, it represents a distorted black hole at merger relaxing to the static Kerr solution. The ring-down phase is normally included in a numerical simulation with the properties compared to the analytic solution as a check.

The search for GWs in detector data requires a large number (${\mathcal O}(10^4)$) of wavform templates to cover the parameter space of a binary merger. It is not practical to generate these templates using numerical relativity since each run takes several weeks.\index{Numerical relativity} Thus, methods have been developed that use the limited numerical relativity results  availableto set certain parameters in formulas that are either algebraic or the solution of an ordinary differential equation, thus enabling the rapid generation of the required waveform templates. The most commonly used such methods are the {\bf Effective One Body (EOB)}\index{Effective One Body (EOB)} approach (see, e.g.,~\cite{Damour2011}), which can be regarded as being mainly an analytical modification of post-Newtonian theory; and {\bf Phenomenological waveform models}\index{Phenomenological waveform models} (e.g.,~\cite{Santamaria2010}), in which the waveform during the merger phase is constructed as a best fit to the numerical relativity data.

An {\bf Extreme Mass Ratio Inspiral (EMRI)}\index{Extreme Mass Ratio Inspiral (EMRI)} is a compact binary inspiral and merger when the mass of one companion is much lower than that of the other~\cite{Poisson11}. Astrophysically, such events occur when a supermassive black hole captures an object with a mass of order a few $M_\odot$. Such events are not in the LIGO/Virgo frequency band, but are expected to be observed by LISA, see Sec. \ref{s-Det}.\index{LIGO}\index{LISA}

The result of GW calculations is often reported in terms of $\psi_4$ (rather than $h_+,h_\times$), and the result decomposed into functions denoted by ${}_sY^{\ell,m}$. What are these quantities? The Newman-Penrose \index{Newman-Penrose quantity $\psi_4$} quantity $\psi_4$ is a complex quantity defined in terms of the Weyl tensor and a null tetrad, and is related to $h_+,h_\times$ by
\begin{equation}
\psi_4=\partial^2_t(h_+-i h_\times)\,.
\end{equation}
The ${}_sY^{\ell,m}$ are spin-weighted spherical harmonics\index{Spin-weighted spherical harmonic ${}_sY^{\ell,m}$} that can be regarded as generalizations of the familiar spherical harmonics $Y^{\ell,m}$ to be able to represent vector and tensor quantities over the unit sphere. In order to describe quadrupolar GWs, only a limited set of the ${}_sY^{\ell,m}$ will be needed, specifically
\begin{align}
{}_2Y^{2,2}&=\frac{\sqrt{5}}{8\sqrt{\pi}}\exp(2i\phi)(1-\cos\theta)^2\,,\;
{}_2Y^{2,-2}=\frac{\sqrt{5}}{8\sqrt{\pi}}\exp(-2i\phi)(1+\cos\theta)^2\,\nonumber \\
{}_{-2}Y^{2,2}&=\frac{\sqrt{5}}{8\sqrt{\pi}}\exp(2i\phi)(1+\cos\theta)^2\,,\;
{}_{-2}Y^{2,-2}=\frac{\sqrt{5}}{8\sqrt{\pi}}\exp(-2i\phi)(1-\cos\theta)^2\,.
\end{align}
Then, for example, Eq.~(\ref{e-hcircbin}), which gives the GW emission of a circular binary, can be re-expressed as\index{Orbiting binary}
\begin{align}
\psi_4&=\frac{32\mu  r_0^2\omega^4\sqrt{5\pi}}{5r} \times\nonumber \\
&\left[
\cos(2\omega t)\left({}_{-2}Y^{2,-2}+{}_{-2}Y^{2,2}\right)
+\sin(2\omega t)\left(i\,{}_{-2}Y^{2,-2}-i\,{}_{-2}Y^{2,2}\right)\right]\,.
\end{align}
We refer to the literature for a fuller discussion of the definition and properties of the Newman-Penrose scalars, and of spin-weighted spherical harmonics, e.g.~\cite{Alcubierre:2008,Baumgarte2010,Bishop16}.

\section{\textit{The gravitational wave field of an orbiting binary}}
\label{s-OrbBin}
\index{Orbiting binary}
\subsection{\textit{Inspiral rate and time to merger}}
\label{s-inspiral}

\begin{figure}
\begin{center}
\includegraphics[width=5cm,angle=270]{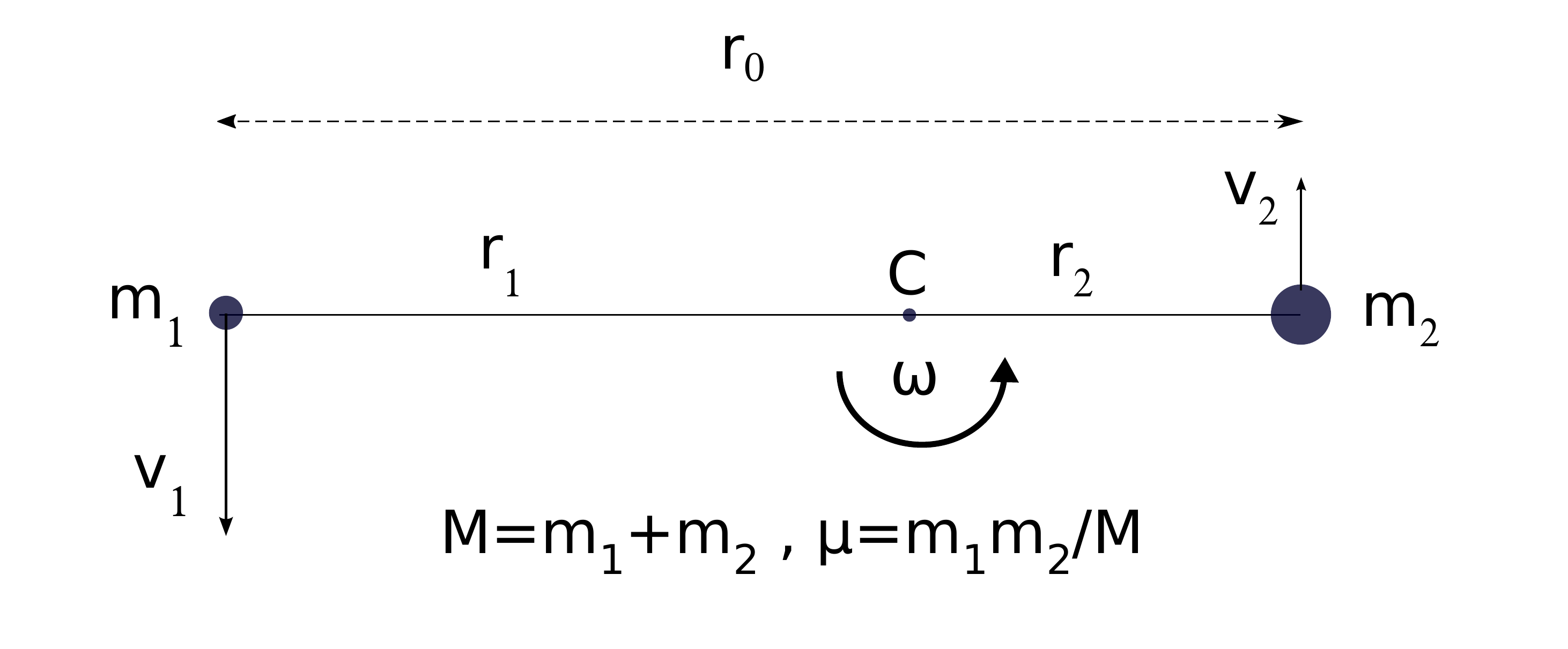}
\end{center}
\caption{Variables used in the orbiting circular binary calculations}
\label{f-cirbin}
\end{figure}

We now consider the long-term behaviour of a binary system in circular orbit, with the GW field given by Eq.~(\ref{e-hcircbin}) and consequently the energy loss $L_{GW}$ given by Eq.~(\ref{e-LGW-circbin}). The variables used are illustrated in Fig.~\ref{f-cirbin}: there are two masses $m_1,m_2$ with velocities $v_1,v_2$ respectively in circular orbit of diameter $r_0$ and angular velocity $\omega$, and let $M=m_1+m_2, \mu=m_1m_2/M$. Let $r_1,r_2$ be the distances from $m_1,m_2$ respectively to the centre of mass of the system $C$ so that $r_1+r_2=r_0$ and $r_1m_1=r_2m_2$; solving these two equations leads to $r_1=r_0 m_2/M,r_2=r_0m_1/M$. In Newtonian theory, gravitational attraction is balanced by the orbital centripetal acceleration, which leads to
\begin{equation}
\omega^2=\frac{M}{r_0^3}\,,
\label{e-omMr0}
\end{equation}
then the Newtonian kinetic plus potential energy of the system is
\begin{equation}
E=-\frac{\mu M}{r_0}+\frac 12 \omega^2 r_0^2\mu=-\frac{\mu M}{r_0}+\frac{\mu M}{2r_0}
=-\frac{\mu M}{2r_0}\,.
\label{e-orbE}
\end{equation}
Now, conservation of energy gives $\partial_t E+ L^{[GW]}=0$ with $L^{[GW]}$ given by Eq.~(\ref{e-LGW-circbin}), so that
\begin{equation}
\partial_t r_0=-\frac{64 M^2\mu}{5r_0^3}\,,
\label{e-ins-d}
\end{equation}
which easily integrates to give a\index{Time to coalescence} time to coalescence $t_c$ (when $r_0(t)=0$)
\begin{equation}
t_c=\frac{5r_0^4}{256 M^2\mu}\,.
\label{e-ins-tc}
\end{equation}

Eqs.~(\ref{e-ins-d}) and (\ref{e-ins-tc}) lead to important results in GW astrophysics. Consider the example of a binary system comprising two $10M_\odot$ black holes in circular orbit such that the time to coalescence $t_c$ is $10^9$ years: what is the initial diameter of the orbit $r_0$? The value of $r_0$ for which $t_c$ is $10^9$ years is astrophysically important, since if the initial separation were larger (by a factor of 2), then $t_c$ would be greater than the age of the universe and so GW emission alone would not cause the binary to coalesce. Inverting Eq.~(\ref{e-ins-tc}) gives
\begin{equation}
r_0=4\left(\frac {10^9 \mbox{years} \times 2000 M_\odot^3 }{5}  \right)^{\frac 14}
=7.5\times 10^6\;\mbox{km}\,,
\end{equation}
where we have used the conversion factors $1\;$year = $0.95\times 10^{13}\;$km, and $1M_\odot=1.48\;$km. By astronomical standards, the distance $7.5\times 10^6\;$km is very small -- it is about 5 solar diameters. Thus, the formation of coalescing black hole binaries must involve astrophysical processes other than GW emission.

The derivation of Eqs.~(\ref{e-ins-d}) and (\ref{e-ins-tc}) assumed that the system is quasi-static, but that is not the case at, and just before, coalescence. From the quadrupole formula the magnitude of a GW scales as $v^2$ (where $v$ is the orbital velocity), and the error is ${\mathcal O}(v^4)$. The problem being considered determines an acceptable magnitude for the error. Suppose that we take $v_{\mbox{\small max}}=0.1$ (so that the error in the GW is ${\mathcal O}(10^{-4})$), and we find that for the equal mass binary example the orbital diameter when $v=v_{\mbox{\small max}}$ is $r_0=50M$. Working out the time to coalescence gives 12s; of course, this figure is not accurate, but if the time to coalescence is $10^9$ years any error occuring during the last few seconds of the inspiral is negligible.

Differentiating Eq.~(\ref{e-omMr0}) gives
\begin{equation}
\frac{\partial_t\omega}{\omega}=-\frac 32 \frac{\partial_t r_0}{r_0}\,,
\end{equation}
and we can then replace $\partial_t r_0,r_0$ in Eq.~(\ref{e-ins-d}) by $\partial_t\omega,\omega$ to obtain  the chirp mass\index{Chirp mass ${\mathcal M}$} formula
\begin{equation}
{\mathcal M}\equiv\frac{(m_1 m_2)^{3/5}}{M^{1/5}}=\mu^{3/5}M^{2/5}
=\left(\frac{5}{96}\pi^{-8/3}f^{-11/3}(\partial_t f)
\right)^{3/5}\,,
\label{e-chirp}
\end{equation}
where $f$ is the wave frequency, related to the orbital angular velocity by
$\omega=\pi f$. Eq.~(\ref{e-chirp}) will be discussed further in Sec.~\ref{s-Detbsp}: the key point is that both $f$ and $\partial_t f$
are determinable entirely from LIGO/Virgo data, and thus so is the chirp mass ${\mathcal M}$.\index{LIGO}
Observed changes to the orbital frequency of the Hulse-Taylor binary
pulsar~\cite{Taylor79}, and other systems, have provided confirmation of this formula since the 1970s.

The decrease in the Newtonian orbital energy of the binary (Eq.~(\ref{e-orbE})) can be regarded as being caused by a radiation reaction force (also called the ``self-force'')\index{Self-force}. A straightforward Newtonian calculation shows that the radiation reaction is of magnitude
\begin{equation}
F=\frac{32}{5}\mu^2 r_0^3\omega^5\,,
\end{equation}
and is applied to each object in the binary in the opposite direction to that of its orbital motion. Since the radiation reaction comprises two equal and opposite forces a distance $r_0$ apart, it imposes a torque of magnitude $r_0 F$ which is equal to the rate of change of angular momentum of the system. Since angular momentum is conserved, the GWs must be carrying angular momentum away at a rate
\begin{equation}
r_0 F=\frac{32}{5}\mu^2 r_0^4\omega^5=\frac{32\mu^2 M^{5/2}}{5 r_0^{7/2}}\,.
\label{e-r0F}
\end{equation}
Now the angular momentum $J$ of the system is
\begin{equation}
J=\omega(m_1r_1^2+m_2r_2^2)=r_0^2\omega\mu=\mu\sqrt{Mr_0}\,,
\end{equation}
so that
\begin{equation}
\partial_t J=\mu \frac{\sqrt{M}\partial_t r_0}{2\sqrt{r_0}}=
  -\frac{32\mu^2 M^{5/2}}{5 r_0^{7/2}}
  \label{e-dtJ}
\end{equation}
which is consistent with Eq.~(\ref{e-r0F}).
The rate at which GWs transport angular momentum may be expressed in a form similar to Eq.~(\ref{e-LGW}), giving the value found in Eq.~(\ref{e-dtJ}) for a circular binary. This also applies to the transport of linear momentum; although it is zero for the binary being considered here, it can be substantial for two black holes with misaligned spins. These results are not discussed here, but are given, for example, in~\cite{Baumgarte2010}.

\subsection{\textit{Eccentricity reduction by gravitational waves}}
\index{Eccentricity}
In general binary orbits are elliptical rather than circular, but an effect of GWs is to circularize an elliptic orbit. Thus close to merger (when the GW emission is strongest and most likely to be detected), binary orbits have a very low eccentricity and can be treated as circular; the only exception would be the unlikely scenario that the binary is located in a sytem containing other massive objects that increase the eccentricity of the binary. The eccentricity reduction formula was first derived in~\cite{Peters:1964}, and here we just outline the calculation.

An elliptical orbit is described by its semi-major axis $a$ and its eccentricity $e$. In Newtonian theory, these are constants of the motion, and here they are treated as slowly varying functions of time. The orbital equation is
\begin{equation}
r=\frac{2 a (1-e^2)}{1+e\cos\phi}\,,
\end{equation}
where $r$ is the distance between the orbiting objects. A circular orbit is recovered on setting $e=0,a=r_0/2$ so that $r=2a=r_0$. The Newtonian energy and Newtonian angular momentum are
\begin{equation}
E=-\frac{\mu M}{4a}\,,\;\;J=\mu \sqrt{2M a (1-e^2)}\,,
\label{e-EJecc}
\end{equation}
and the rates of energy and angular momentum emission by GWs are
\begin{align}
\left<L^{[GW]}\right>&= -\frac{\mu^2M^3}{5a^5(1-e^2)^{7/2}}\left(1+\frac{73e^2}{24}+\frac{37e^4}{96} \right)\,,\\
\left<\frac{dJ}{dt}\right>&=-\frac{2\sqrt{2}\mu^2 M^{5/2}}{5a^{7/2}(1-e^2)^2}
\left(1+\frac{7e^2}{8}\right)\,,
\label{e-LJtecc}
\end{align}
where$\left< \;\right>$ denotes an average over one orbit. The above formulas reduce to those for a binary in circular orbit on seting $e=0,a=r_0/2$. Substituting Eqs.~(\ref{e-EJecc}) into Eqs.~(\ref{e-LJtecc}) leads to equations for $\partial_t a,\partial_t e$, which are combined to give
\begin{equation}
\left<\frac{da}{de}\right>=\frac{12a}{19e}\left(1+{\mathcal O}(e^2)\right)\,,
\end{equation}
from which it follows that
\begin{equation}
e\approx e_0\left(\frac{a}{a_0}\right)^{19/12}\,.
\end{equation}
During the time that GW emission starts to drive an inspiral until the GWs become detectable, the size of the orbit (i.e. $a$) decreases by a factor of about $10^4$, so that $e\approx 10^{-6} e_0< 10^{-6}$.

\section{\textit{Detection of gravitational waves}}
\label{s-Det}

GW detectors are sensitive only in a specific frequency range, and so key factors for the detection of a GW are its magnitude and frequency, and it is useful to have simple order of magnitude estimates of these quantities. From Eqs.~(\ref{e-hcircbin}) and (\ref{e-omMr0}), it follows that
\begin{equation}
|h_+,h_\times|={\mathcal O}\left(\frac{2\mu M}{r_0 r}  \right)\,,
\end{equation}
which is maximized just before merger when $r_0\sim 2M$ so that
\begin{equation}
|h_+,h_\times|={\mathcal O}\left(\frac{\mu}{r}\right) \approx 0.5\times 10^{-19}
\frac{\mu/1M_\odot}{r/1\mbox{Mpc}} \,.
\end{equation}
The GW frequency is estimated from Eq.~(\ref{e-omMr0}) as $\omega=\sqrt{2}/(4M)$ so that
\begin{equation}
f\sim \frac{\sqrt{2}}{4\pi M}\approx 2\times 10^4\frac{M_\odot}{M}\,\mbox{Hz}\,.
\end{equation}\index{Orbiting binary}
The GW from a compact binary inspiral slowly increases in magnitude and frequency until reaching the maximum values indicated above.

\begin{figure}
\begin{center}
\includegraphics[width=12cm]{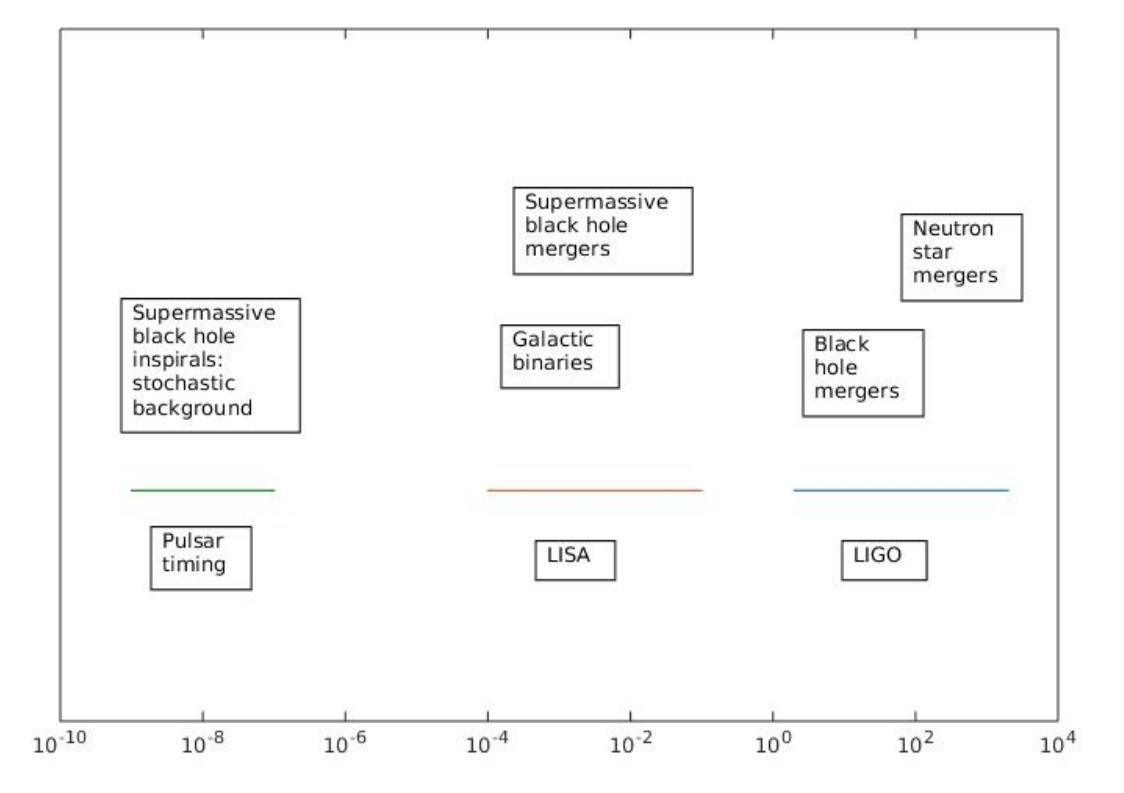}
\end{center}
\caption{LIGO, LISA and Pulsar Timing frequency ranges and target astrophysical
events}
\label{f-det-freq}\index{LIGO}
\end{figure}

The various options for measuring GWs are described in detail in chapters 2 to 7 in this book, and here only an outline is presented.
The frequency ranges in which detection systems are sensitive, and the corresponding astrophysical processes, are summarized in Fig. \ref{f-det-freq}.

\subsection{\textit{GW detection facilities}}
\begin{figure}
\begin{center}
\includegraphics[width=12cm]{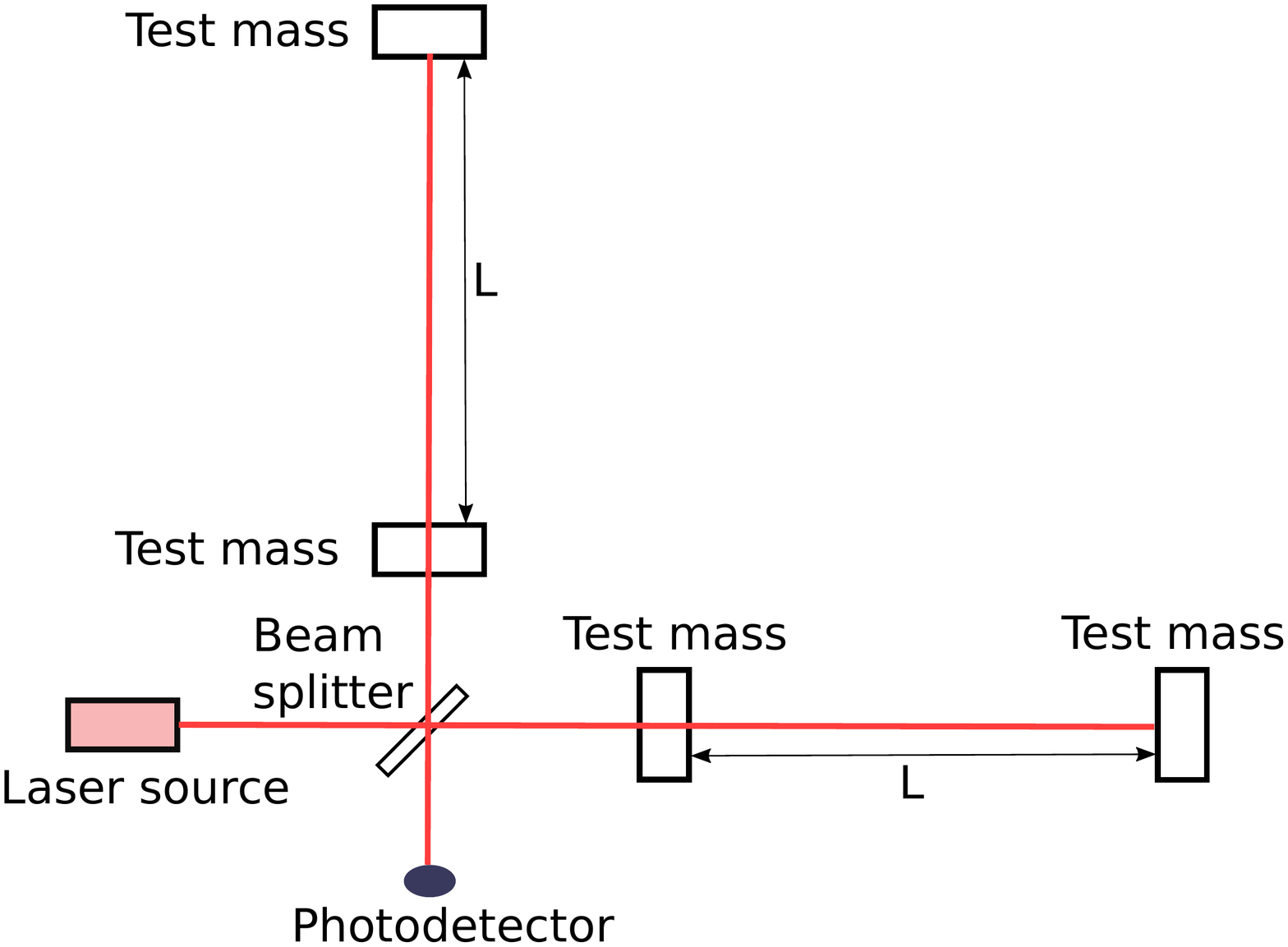}
\end{center}
\caption{Schematic representation of a terrestrial laser interferometer detector. The arm length $L$ is 4km for LIGO and 3km for Virgo.}
\label{f-LIGO}\index{LIGO}
\end{figure}

Ground-based laser interferometer detectors\index{Laser interferometer} detector comprise two perpendicular arms each of length several km, and laser light travels along each arm and back again, see Fig.~\ref{f-LIGO}; then the difference in travel time between the two arms is measured as an interference fringe. These detectors are sensitive in the frequency range $\approx 20$Hz to $\approx 2000$Hz, and have observed compact binary mergers (with the compact objects being neutron stars or stellar-mass black holes) since September 2015~\cite{Abbott2016a}. Other potential GW sources that may be detected are discussed in Sec.~\ref{s-obs}. Detector sensitivity has been continually improved since the first science run in 2001. This is important, since a 10-fold increase in sensitivity means a 10-fold increase in the distance at which a source may be detected (see Eq.~\ref{e-quad3}), which increases the volume searched and therefore the expected detection rate by a factor of $1000$. Currently the network comprises operational detectors GEO600\index{GEO600} in Germany (this is the least sensitive detector), KAGRA\index{KAGRA} in Japan, LIGO Livingstone in USA, LIGO Hanford in USA and VIRGO in Italy, with LIGO-India scheduled to be operational in 2024.\index{LIGO}\index{Virgo} In order for there to be a high probability that an observation of a GW event is real rather than an instrument glitch, it is important that it be observed in at least two detectors. Further, the accuracy of sky localization increases with the number of (well-separated) detectors in which the event is observed; accurate sky localization improves the chance that an EM counterpart can be identified, leading to a fuller picture of the astrophysics of the event.

The Laser Interferometer Space Antenna (LISA)\index{LISA} is a mission of the European Space Agency and is scheduled for launch in 2034. It comprises three satellites at the vertices of an equilateral triangle with sides $2.5\times 10^6$km in an Earth-like orbit around the Sun. It will be sensitive to GWs in the frequency band $10^{-5}$Hz to $1$Hz. The GW events that are expected to be observed include the merger of supermassive black holes and EMRIs.\index{Extreme Mass Ratio Inspiral (EMRI)} LISA should also detect GWs from close binaries comprising stellar-mass black holes, neutron stars or white dwarfs, and should also be able to predict the end-point of an inspiral, i.e. the time of a compact binary merger event detectable by a gound based facility.

Millisecond pulsars can be regarded as highly accurate clocks. 
Pulsar timing\index{Pulsar timing} projects
measure accurately the time of arrival of each pulse from a particular pulsar, and compare it to the
expected time of arrival assuming a uniform time between the pulses. This difference is recorded
as the timing residual. (In practice, the process is rather more complicated, because it is only when
averaged over a number of pulses that uniform pulse emission occurs, and further there may be
glitches in pulse emission; thus the data analysis is statistical rather than direct). These measurements are repeated for a number of pulsars distributed over the sky. If a GW passes
the Earth, then the timing residuals will oscillate according to the form of the GW. Pulsar timing is sensitive to GWs at very low frequencies, about
$10^{-7}$Hz to $10^{-9}$Hz corresponding to periods of a few months up to a decade, generated, for example, by super-massive black hole binaries well
before merger. Data collection has been ongoing since 2004, but to date has only
been able to set upper limits.

\subsection{\textit{Data interpretation and determination of source parameters}}
\label{s-Detbsp}
\index{Orbiting binary}\index{Source parameter determination}
A major aspect of laser interferometer data analysis is to identify GW signals that are hidden by noisy data. The methods used vary according to the type of signal being searched for; these matters are not pursued here, but are discussed in detail in chapters 30 to 36. The presentation here is limited to an outline of the determination of the astrophysical parameters from GW data of a binary black hole merger.

Suppose that a GW signal is identified in detectors $A$ and $B$, and that it is observed in detector $B$ a time $t_{AB}$ after it is observed in detector $A$. Then if $\Delta_{AB}$ is the light travel time from $A$ to $B$, the direction to the GW source must be at an angle
\begin{equation}
\alpha=\arccos\left(\frac{t_{AB}}{\Delta_{AB}}\right)
\end{equation}
to the line $AB$. In practice, there is an error involved in the measurement of $t_{AB}$, which means that the source is located within an annulus, rather than on a circle, on the sky. Further, as well as the timing of the GW signal, we also know its magnitude in each detector, and when combined with a model of the source this may lead to a further restriction on the direction to the source; i.e., the annulus may be reduced to a partial annulus. In the case that the GW event is seen in 3 or more detectors, the sky localization reduces to a point if there are no measurement errors, and in practice to a small region of the sky.

\begin{figure}
\begin{center}
\includegraphics[width=12cm]{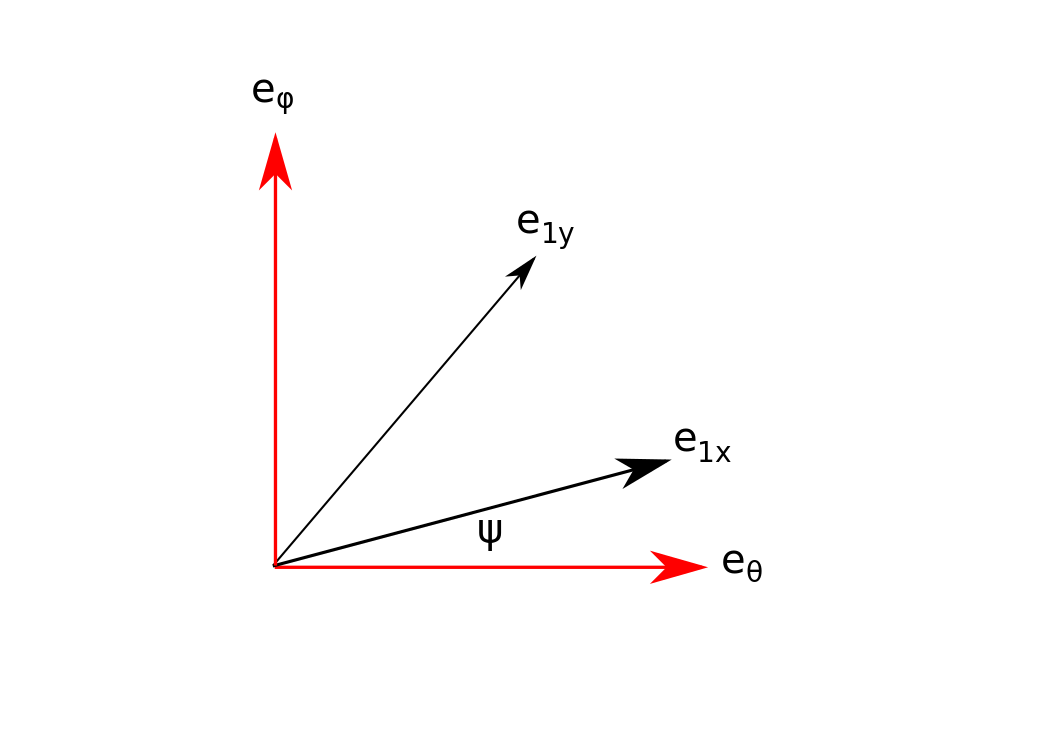}
\end{center}
\caption{Detector arms projected into plane N}
\label{f-PlaneN}
\end{figure}
\index{Quadrupole formula}
For the next steps, we assume that the quadrupolar formulas remain valid. The chirp mass ${\mathcal M}$ \index{Chirp mass ${\mathcal M}$}is determined using Eq.~(\ref{e-chirp}). Then, using Eq.~(\ref{e-omMr0}) and $\omega=\pi f$ in Eq.~(\ref{e-hcircbin}) gives
\begin{align}
(h_+,h_\times)&=-\frac{2{\mathcal M}^{5/3}(\pi f)^{2/3}}{r}\nonumber \\
&\times\left((1+\cos^2\theta)\cos(2\omega(t-r)-2\phi),2\cos\theta\sin(2\omega(t-r)-2\phi)\right) \,.
\label{e-hcircbinchirp}
\end{align}
Thus, a measurement of the polarization modes $h_+,h_\times$\index{Polarization modes $h_+,h_\times$} would lead to the ratio  $h_\times/h_+$ which can be solved for the angle $\theta$ (which is the polar coordinate of the Earth in the reference frame of the source), and then Eq.~(\ref{e-hcircbinchirp}) may be solved to find the distance $r$ between source and Earth. If the event is well-localized on the sky, then the plane $N$ transverse to the propagation direction is known. Within $N$, let $\bm{e}_\theta,\bm{e}_\phi$ be unit vectors in the $\theta,\phi$ directions of the source frame, with respect to which $h_+,h_\times$ of Eq~(\ref{e-hcircbinchirp}) are defined. Let $\bm{e}_{1x},\bm{e}_{1y}$ be unit vectors representing the directions of the arms of a detector $D_1$ projected into $N$, and let the angle between $\bm{e}_\theta$ and $\bm{e}_{1x}$ be $\psi$. The angle between $\bm{e}_{1x}$ and $\bm{e}_{1y}$ is known, so that between $\bm{e}_\theta$ and $\bm{e}_{1y}$ is known in terms of $\psi$. These quantities are illustrated in Fig.~\ref{f-PlaneN}. Now, given $h_+,h_\times$ and $\psi$, we can use Eq.~(\ref{e-rotate}) to rotate the coordinate axes and find the change in length of each arm of $D_1$ due to the GW, and the difference between these changes is the detector response. Thus the measured detector response leads to one equation involving the three unknowns $h_+,h_\times$ and $\psi$. Measurements in detectors $D_2,D_3$ (and if possible, more detectors so as to reduce error bars) then provide sufficient information to close the system and solve for the unknowns.

The remaining astrophysical parameters are determined using features of the waveform beyond the quadrupole formula. A template bank of merger waveforms over the parameter space
to be explored is needed. For black hole mergers, the frequency of the GWs scales as the inverse of the total mass $M$, and the magnitude scales as $M$; then the parameter space comprises the mass ratio
$q=m_1/m_2$, and the spins $\bm{S}_1, \bm{S}_2$. The mismatch between the calculated waveform ($h_c$) and the observed waveform ($h_o$)\index{Mismatch between waveforms}
is defined as
\begin{equation}
{\mathcal M}(h_c)= 1- \langle h_c,h_o \rangle
\end{equation}
where the inner product is evaluated in the Fourier domain, and includes the detector response function of each detector in which a signal is observed. It is normalized so that it satisfies the conditions $0\le \langle h_1,h_2 \rangle \le 1$
with $\langle h_1,h_2 \rangle =1$ iff $h_1=$ constant $\times h_2$. Then ${\mathcal M}(h_c)$ is minimized
over the template bank. Since $h_o$ is subject to measurement error, the values found for the various astrophysical parameters are also subject to error.

\subsection{\textit{GW observations: Contributions to Physics and Astrophysics}}
\label{s-obs}

\subsubsection{\textit{Fundamental physics}}
\index{LIGO}\index{Virgo}
The inspiral and merger of a binary neutron star system was observed by LIGO/Virgo as GW170817~\cite{Abbott2017f,Abbott2017c}. It was followed, $1.7$s after merger in the GW signal, by a gamma-ray burst GRB 170817A. The distance to the system is $40$Mpc, so we can deduce that the speed of GWs ($c_{\tiny{\mbox{GW}}}$) is very close to being exactly that of light ($c_{\tiny{\mbox{EM}}}$) \index{Speed of gravitational waves}~\cite{Abbott2017GRB}
\begin{equation}
1-3\times 10^{-15}\le \frac{c_{\tiny{\mbox{GW}}}}{c_{\tiny{\mbox{EM}}}}\le 1+7\times 10^{-16}\,.
\end{equation}
In general relativity, the speed of GW propagation is exactly the same as that of light so this result is a strong test of the theory. Further, all GW waveforms observed to date have been consistent with the predictions of general relativity. These waveforms have been produced by compact object mergers and include the strong field regime. (Previously, experiments have only been able to test the theory in the regime of small deviations from Newtonian theory). Again, this amounts to a strong test of general relativity.

\subsubsection{\textit{Cosmology}}
\index{Orbiting binary}
Data for the binary neutron star event GW170817 was taken by three detectors enabling the angle $\theta$ to be constrained; then Eq.~(\ref{e-hcircbinchirp}) is used to provide an independent distance estimate. This is an example of a compact binary merger being a standard siren~\cite{Schutz86b}. Since an optical counterpart was identified, the red-shift of the system is known to good precision, and thus the expansion rate, i.e. the Hubble constant, can be measured. The result is $70.0^{+12.0}_{-8.0}$km/s/Mpc\index{Hubble constant}. The
uncertainty in the value should be reduced in future as more events are discovered. The value obtained is consistent with estimates from both cosmic microwave background data and supernovae (SN1a), and does not yet shed light on the tension between these values.

\subsubsection{\textit{Astrophysics}}
The first GW detection, GW150914\cite{Abbott2016a}, was the merger of two black holes of mass approximately $36M_\odot$ and $29M_\odot$ to give a $62M_\odot$ black hole. Until then, the most massive black hole that had been observed was $\approx 25M_\odot$. As more and more such events are detected, the mass distribution of black holes in the local universe, as well as the rate at which mergers occur, will be well-constrained, and this information will enable improved modeling of black hole, and black hole merger, formation channels. Similar remarks apply to neutron star mergers, although to date only two such events have been reported. Event GW190814~\cite{Abbott2020} was particularly interesting because it involved the merger of a $23M_\odot$ black hole with a $2.6M_\odot$ compact object, which could be either the most massive neutron star, or the smallest black hole, to have been observed.

Besides compact object mergers, current detectors are expected to observe GWs from supernovae, as well as the continuous signal from any asymmetry in a millisecond pulsar (which is a rapidly rotating neutron star). Such signals have not yet been detected, but their observation should lead to significant improvement to the modeling of these systems.

\section{\textit{Conclusion}}
\label{s-Conc}
GWs in general relativity were first predicted in 1916, but it was nearly 100 years until the first direct detection of a GW event in 2015. Detections of new events are becoming more and more frequent, and the results to date have already had an impact on astrophysics, cosmology and physics. This impact will grow as detector sensitivities are improved and new facilities come online, so that more events, and of different types, are detected. The future of GW astronomy is, of course, unknown, but we can expect that it will bring some fascinating results, and probably some surprises.


\bibliography{aeireferences,addref}

\begin{thebibliography}{10}

\bibitem{Einstein1916}
A.~Einstein, ``Naherungsweise {I}ntegration der {F}eldgleichungen der
  {G}ravitation,'' {\em Sitzungsberichte der K{\"o}niglich Preussischen
  Akademie der Wissenschaften (Berlin)}, vol.~1916, pp.~688--696, 1916.

\bibitem{Einstein1918}
A.~Einstein, ``{\"U}ber {G}ravitationswellen,'' {\em Sitzungsberichte der
  K{\"o}niglich Preussischen Akademie der Wissenschaften (Berlin)}, vol.~1918,
  pp.~154--167, 1918.

\bibitem{Kennefick:1997kb}
D.~Kennefick, ``{Controversies in the History of the Radiation Reaction problem
  in General Relativity},'' 1997.

\bibitem{Bondi62}
H.~Bondi, M.~G.~J. van~der Burg, and A.~W.~K. Metzner, ``Gravitational waves in
  general relativity {VII}. {W}aves from axi-symmetric isolated systems,'' {\em
  Proc. R. Soc. London}, vol.~A269, pp.~21--52, 1962.

\bibitem{Sachs62}
R.~Sachs, ``Gravitational waves in general relativity {VIII}. {W}aves in
  asymptotically flat space-time,'' {\em Proc. Roy. Soc. London}, vol.~A270,
  pp.~103--126, 1962.

\bibitem{Isaacson68}
R.~Isaacson, ``Gravitational radiation in the limit of high frequency. {II}.
  nonlinear terms and the effective stress tensor,'' {\em Phys. Rev.},
  vol.~166, pp.~1272--1280, 1968.

\bibitem{Peters:1963ux}
P.~C. Peters and J.~Mathews, ``Gravitational radiation from point masses in a
  {K}eplerian orbit,'' {\em Phys. Rev.}, vol.~131, pp.~435--440, 1963.

\bibitem{Peters:1964}
P.~C. Peters, ``Gravitational radiation and the motion of two point masses,''
  {\em Phys. Rev.}, vol.~136, pp.~B1224--B1232, 1964.

\bibitem{Weber:1969bz}
J.~Weber, ``{Evidence for discovery of gravitational radiation},'' {\em Phys.
  Rev. Lett.}, vol.~22, pp.~1320--1324, 1969.

\bibitem{Taylor79}
J.~Taylor, L.~Fowler, and P.~McCulloch, ``Measurements of general relativistic
  effects in the binary pulsar psr1913 + 16,'' {\em Nature}, vol.~277,
  pp.~437--440, 1979.

\bibitem{Lorentz17}
H.~Lorentz and J.~Droste, ``The motion of a system of bodies under the
  influence of their mutual attraction, according to {E}instein’s theory,''
  {\em Versl. K. Akad. Wet. Amsterdam}, vol.~26, pp.~392, 649, 1917.

\bibitem{Lorentz37}
H.~Lorentz and J.~Droste, ``The motion of a system of bodies under the
  influence of their mutual attraction, according to {E}instein’s theory,''
  in {\em The Collected Papers of H.A. Lorentz, Vol. 5,} (H.~Lorentz, ed.),
  pp.~330--355, The Hague, Netherlands: Nijhoff, 1937.

\bibitem{Arnowitt62}
R.~{Arnowitt}, S.~{Deser}, and C.~W. {Misner}, ``{Republication of: The
  dynamics of general relativity},'' {\em General Relativity and Gravitation},
  vol.~40, pp.~1997--2027, Sept. 2008.

\bibitem{Smarr77}
L.~Smarr, ``Spacetimes generated by computers: Black holes with gravitational
  radiation,'' {\em Ann. N. Y. Acad. Sci.}, vol.~302, pp.~569--604, 1977.

\bibitem{Pretorius:2005gq}
F.~Pretorius, ``Evolution of binary black hole spacetimes,'' {\em Phys. Rev.
  Lett.}, vol.~95, p.~121101, 2005.

\bibitem{Abbott2016a}
{B. P. Abbott et. al.}, ``{Observation of Gravitational Waves from a Binary
  Black Hole Merger},'' {\em Phys. Rev. Lett.}, vol.~116, p.~061102, 2016.

\bibitem{Alcubierre:2008}
M.~Alcubierre, {\em Introduction to $3+1$ {N}umerical {R}elativity}.
\newblock Oxford, UK: Oxford University Press, 2008.

\bibitem{Blanchet06}
L.~Blanchet, ``Gravitational radiation from post-{N}ewtonian sources and
  inspiralling compact binaries,'' {\em Living Rev. Relativ.}, vol.~9, p.~4,
  2006.

\bibitem{Baumgarte2010}
T.~W. {Baumgarte} and S.~L. {Shapiro}, {\em {Numerical Relativity: Solving
  Einstein's Equations on the Computer}}.
\newblock Cambridge University Press, Cambridge UK, 2010.

\bibitem{Chandrasekhar98}
S.~Chandrasekhar, {\em {The mathematical theory of black holes}}.
\newblock New York: Oxford University Press., 1998.

\bibitem{Kokkotas99a}
K.~{Kokkotas} and B.~{Schmidt}, ``{Quasi-Normal Modes of Stars and Black
  Holes},'' {\em Living Reviews in Relativity}, vol.~2, p.~2, Sept. 1999.

\bibitem{Damour2011}
T.~Damour and A.~Nagar, {\em The Effective One-Body Description of the Two-Body
  Problem}, pp.~211--252.
\newblock Dordrecht: Springer Netherlands, 2011.

\bibitem{Santamaria2010}
L.~{Santamar{\'{\i}}a} {\em et~al.}, ``{Matching post-Newtonian and numerical
  relativity waveforms: Systematic errors and a new phenomenological model for
  nonprecessing black hole binaries},'' {\em Phys. Rev. D}, vol.~82, p.~064016,
  Sept. 2010.

\bibitem{Poisson11}
E.~Poisson, A.~Pound, and I.~Vega, ``The motion of point particles in curved
  spacetime,'' {\em Living Reviews in Relativity}, vol.~14, 2011.

\bibitem{Bishop16}
N.~T. Bishop and L.~Rezzolla, ``Extraction of gravitational waves in numerical
  relativity,'' {\em Living Reviews in Relativity}, vol.~19, 2016.

\bibitem{Abbott2017f}
{B. P. Abbott et. al.}, ``{GW170817: Observation of Gravitational Waves from a
  Binary Neutron Star Inspiral},'' {\em Phys. Rev. Lett.}, vol.~119, p.~161101,
  2017.

\bibitem{Abbott2017c}
{B. P. Abbott et. al.}, ``{Multi-messenger Observations of a Binary Neutron
  Star Merger},'' {\em Astrophys. J. Lett.}, vol.~848, p.~L12, 2017.

\bibitem{Abbott2017GRB}
{B. P. Abbott et. al.}, ``{Gravitational Waves and Gamma-Rays from a Binary
  Neutron Star Merger: GW170817 and GRB 170817A},'' {\em Astrophys. J. Lett.},
  vol.~848, p.~L13, 2017.

\bibitem{Schutz86b}
B.~F. Schutz, ``Determining the hubble constant from gravitational wave
  observations,'' {\em Nature}, vol.~323, pp.~310--311, 1986.

\bibitem{Abbott2020}
{B. P. Abbott et. al.}, ``{GW190814: Gravitational Waves from the Coalescence
  of a 23 Solar Mass Black Hole with a 2.6 Solar Mass Compact Object},'' {\em
  Astrophys. J. Lett.}, vol.~896, p.~L44, 2020.

\end{thebibliography}

\input{bishop.ind}

\end{document}